\pdfoutput=1
\documentclass[12pt]{article}

\usepackage[a4paper, margin=2.5cm]{geometry}
\usepackage{amsmath, amssymb, amsfonts, bm}
\usepackage{graphicx}
\usepackage{booktabs}
\usepackage{array}  \usepackage{natbib}
\usepackage{setspace}
\usepackage{hyperref}
\usepackage{xcolor}
\usepackage{float}
\usepackage{subcaption}
\usepackage{algorithm}
\usepackage{algpseudocode}
\usepackage{tikz}
\usepackage{enumitem}
\usepackage{appendix}
\usepackage{longtable}
\usepackage{sectsty}

\newcommand{\R}{\mathbb{R}}

\newcommand{\bx}{\bm{x}}

\definecolor{MyDarkBlue}{RGB}{0, 0, 139}    
\definecolor{MyDarkGreen}{RGB}{0, 100, 0}   
\definecolor{MySectionBlue}{RGB}{0, 50, 200} 
\definecolor{MyURLred}{RGB}{153, 0, 0}      

\sectionfont{\color{MyDarkBlue}}
\subsectionfont{\color{MyDarkGreen}}
\hypersetup{
    colorlinks=true,
    linkcolor=blue!60!black,
    citecolor=MyURLred,
    urlcolor=blue!60!black
}

\title{\textcolor{MyDarkBlue}{\textbf{Deep-$\pi^*$: A Neural Network Estimate of Trend Inflation}}\thanks{I am grateful to Jes\'us Fern\'andez-Villaverde, to several members of the Banco de Espa\~{n}a and to the participants in the ECB's seminars for useful discussions. All remaining errors are my own.}}

\author{Carlos Montes-Gald\'{o}n\thanks{European Central Bank, Directorate General Economics. Email: \href{mailto:carlos.montes-galdon@ecb.europa.eu}{Carlos.Montes-Galdon@ecb.europa.eu}}}

\date{October 2026}

\begin{document}

\maketitle

\begin{abstract}
\noindent Most empirical models of trend inflation treat it as a latent random walk driven by unobserved shocks, so they cannot say which economic conditions move it. Deep-$\pi^*$ makes the innovations to the trend a possibly nonlinear function of a large number of macroeconomic indicators, estimated with neural networks, so that Shapley values can be used to attribute every movement of the trend to the indicators behind it. Two complementary loss functions define a current trend, which fits inflation today, and a forward-looking trend, which best predicts inflation over the next year. In US data, employment-related variables were the largest contributors to the post-pandemic rise of the current trend. However, the forward-looking trend rose earlier, with real activity, employment, and money and credit as the main drivers, and peaked eight to twelve months before the current trend. Both trends looked through most of the movements in energy and commodity prices. The contributions of the indicators are nonlinear, with sizeable interactions between them. Re-estimated every month during the surge, the model forecast inflation more accurately than the standard UC-SV model and a random walk, and in early 2021, when professional forecasters expected inflation of 3\% or less, the forward-looking trend was already signalling an increase in persistent inflation.

\vspace{0.5cm}
\noindent \textbf{Keywords:} Trend inflation, neural networks, attention mechanism, unobserved components, interpretable machine learning, Shapley values, FRED-MD \\
\textbf{JEL Classification:} C45, C53, E31, E37, E52
\end{abstract}

\vfill
\noindent\fbox{\parbox{\dimexpr\textwidth-2\fboxsep-2\fboxrule\relax}{\small \textbf{Disclaimer:} This paper should not be reported as representing the views of the European Central Bank (ECB). The views expressed are those of the author and do not necessarily reflect those of the ECB or the Eurosystem.}}

\newpage

\tableofcontents

\newpage

\section{Introduction}
\label{sec:introduction}

Understanding persistent movements in inflation is central to the conduct of monetary policy. Central banks around the world, including the Federal Reserve and the European Central Bank, anchor their policy decisions on assessments of where inflation is heading in the medium term, which depends critically on distinguishing short-lived transitory price fluctuations from movements in the underlying trend or more persistent shocks to inflation. However, as \citet{StockWatson2016} emphasise, ``because there are multiple sources of noise in inflation data and because the nature of the noise can change over time, the task of estimating trend inflation is both difficult and of ongoing relevance.'' The post-pandemic inflation episode, in which consumer prices surged to levels unseen in four decades before gradually receding, has made this challenge particularly acute, forcing policymakers to evaluate in real time whether the inflationary shock was cyclical or structural.

Yet the concept of ``trend inflation'' admits several definitions. The trend can be defined as the persistent component of current inflation, as the best predictor of inflation over the coming years, or as long-run inflation expectations. As \citet{FaustWright2013} emphasise in their survey of inflation forecasting, accurate forecasts of inflation must account for its slowly varying local mean, which can be measured, for example, by long-run survey expectations or by the permanent component of an unobserved components model. The choice of definition became especially consequential during 2021--2023, when policymakers at the Federal Reserve and the European Central Bank needed to assess, in real time, whether the post-pandemic price surge was fundamentally transitory or was becoming embedded in the economy's underlying inflation dynamics---a distinction that mattered for the pace and magnitude of monetary tightening, since large surges of inflation have historically been persistent \citep{BlancoEtAl2022}.

The dominant approach to estimating trend inflation relies on unobserved components (UC) models in the tradition of \citet{StockWatson2007}, who propose decomposing inflation into a persistent trend component and a serially uncorrelated transitory component, with time-varying volatility in both. In these models, the trend typically follows a random walk or near-random-walk process, so that changes in trend inflation are driven entirely by unpredictable shocks. Some extensions have enriched this basic framework: \citet{CogleyPrimiceriSargent2010} allow for time-varying persistence in the inflation gap using VAR models, \citet{ChanClarkKoop2018} use long-run inflation expectations from surveys as a signal of the trend, and \citet{ClarkDoh2014} systematically compare alternative trend specifications. However, these models have two common limitations. First, movements in the trend are by construction unexplainable---they cannot be attributed to observable economic forces, which limits the model's usefulness for policy narrative and communication. When a policymaker asks ``why did trend inflation rise?'', a random-walk model can only answer ``because of an unpredictable shock.'' Second, the linear, Gaussian framework may be inadequate to capture the complex, potentially nonlinear relationships between macroeconomic fundamentals and persistent inflation dynamics, especially in periods in which the link between economic slack and inflation has been debated, as in the Great Recession \citep{BallMazumder2011, DelNegroGiannoniSchorfheide2015}. 

This paper proposes \textbf{Deep-$\pi^*$}, a model that addresses both limitations by making the innovations to the trend a function of observable macroeconomic conditions, estimated with neural networks. The trend, $\pi^*$, evolves as a highly persistent autoregressive process:
\begin{align*}
    \pi_t &= \sum_{p=1}^P \phi_p \pi_{t-p} + \left(1 - \sum_{p=1}^P \phi_p\right) \pi_t^* + \sigma_t \varepsilon_t, \\
    \pi_t^* &= \theta \, \pi_{t-1}^* + (1-\theta) \, \tau_t,
\end{align*}
where $\tau_t$ is the innovation to the trend and $\sigma_t$ is the scale of inflation surprises. A key novelty is that we assume that the innovation is a possibly nonlinear function of a large vector of macroeconomic indicators, $\bx_t \in \R^k$, which we approximate with neural networks. 

We should emphasise that the trend in Deep-$\pi^*$ has no stochastic shock of its own. The innovation $\tau_t$ is a deterministic function of the indicators, so that, given the parameters of the network and the initial value of the trend, the indicators observed up to month $t$ determine the trend in month $t$. This is the main difference from UC models, in which the trend is a latent state driven by its own unobserved shock and has to be inferred from the data by filtering, and it has two consequences. First, the likelihood is available in closed form, and the model is estimated by gradient descent, like other neural networks, without a Kalman or particle filter. Second, the uncertainty around the trend comes from the estimation of the network, which we measure with Monte Carlo dropout and with an ensemble of networks trained from different initialisations.

To estimate the model, and in particular the function that maps the indicators to the innovations, we consider four different neural network architectures, which combine two choices. The first choice is about how the different indicators influence the innovation to the trend. We consider first an \textit{encoder} network which progressively compresses the indicators of each month into a scalar innovation to the trend, which can be seen as a nonlinear analogue of principal component analysis. Alternatively, we consider a \textit{similarity-based causal attention} applied before the same encoder which constructs, for each month, a weighted average of the indicators of the current and past months. The weights grow with the cosine similarity between the macroeconomic conditions of the two months and decay geometrically with the distance between them. This second option brings information from past values of the indicators into the trend without fixing an arbitrary number of lags or significantly increasing the number of parameters in the model. The data then determine how far back the attention looks and how much the similarity of conditions matters. 

The second choice is how to model the scale of inflation surprises, $\sigma_t$, which is allowed to be constant or time-varying. The time-varying scale follows a \textit{volatility state} with a recursion similar to that of the trend, and a second, small network maps the indicators and the recent errors of the inflation equation into the innovations of the state. The volatility state is therefore the observation-driven counterpart of the stochastic volatility process used in models of trend inflation: its likelihood is available in closed form, as for the trend, and it is estimated jointly with it. In periods of large surprises, the data then carry less weight in the estimation of the trend, and the predictive densities of inflation widen.

We also propose to estimate the model with two loss functions, which define two different, although complementary, concepts of trend inflation. The first, which we call the \textit{Econometrician's loss}, is the negative log-likelihood of current inflation under a Student-$t$ distribution. It yields the \textit{current trend}: the persistent component of inflation today, given the information available today. The second, the \textit{Central Banker's loss}, is the negative log predictive density of the model's forecasts of inflation over the next 12 months. It yields the \textit{forward-looking trend}: the trend inflation today that gives the most accurate forecasts of inflation over the next year, made with the information up to today, the concept most relevant for a policymaker who sets policy for the medium term. The two trends answer different questions, and we report both, for all four architectures. Comparing what drove them during the post-pandemic surge separates the indicators that carried information about where inflation was going from those that explained where it was at the time. Moreover, because each loss defines its own decision problem, the architecture that serves one loss best need not serve the other, and we therefore compare the architectures separately under each loss.

We then apply the model to US CPI inflation from February 1970 to December 2025, using 122 macroeconomic indicators from the FRED-MD database \citep{McCrackenNg2016}. The eight estimates agree on the main episodes of US inflation history. With time-varying volatility, the trend peaks at about 13\% in 1979--80, averages about 3\% during the Great Moderation and about 2\% in the 2010s, and rises to between 6.7\% and 7.5\% after the pandemic, its highest level since 1981. The two architectures give similar trends, and adding time-varying volatility changes them little. The two losses, by contrast, give trends with the same low-frequency movements but a different timing: in the full-sample estimates, the forward-looking trend peaked in March 2021, eight to twelve months before the current trend. Finally, the volatility of inflation surprises is high in the mid-1970s, in the early 1980s, in 2008 and in 2021--22, and lowest in the 1990s, and it follows the stochastic volatility of the unobserved components model with stochastic volatility (UC-SV) of \citet{StockWatson2007} closely while moving more smoothly.

A key feature of Deep-$\pi^*$ relative to UC models of trend inflation is that it is interpretable by construction. Because the trend has no shock of its own, each of its movements can be traced back to the macroeconomic indicators that we use in the model. We form nine groups of indicators and attribute the innovation of each month to them with Shapley values \citep{Shapley1953}. 

We use the Shapley-based decompositions to show what drove the post-pandemic rise in trend inflation. The labour market drove the rise of the current trend in both architectures: employment is the largest contributor to its rise from January 2019 to its peak, with 41 to 52\% of it. This group includes payroll employment and hours, the unemployment rate and claims, and the help-wanted indices, which measure job vacancies. Payroll employment and the help-wanted indices add to the rise in both architectures. The result is in line with the discussion at the time, in which the tight labour market was seen as a source of inflation \citep{DomashSummers2022, BallLeighMishra2022, BenignoEggertsson2023}. It is also consistent with \citet{BernankeBlanchard2025}, who do not study trend inflation but find that, in the United States, the effects of a tight labour market on inflation are more persistent than those of shocks to product prices, while in the euro area the same model finds a small contribution of labour market tightness to wage growth and inflation \citep{ArceEtAl2024}. For the forward-looking trend, the two architectures attribute the rise differently: the encoder attributes most of it to real activity and employment, and the attention architecture mainly to the surge of money and credit in 2020.

These drivers also speak to the distinction between demand and supply. After the pandemic, the trends moved mainly with indicators of demand and of monetary conditions, and they looked through most of the movements in prices, such as those of energy, that are usually associated with supply shocks and whose effects on inflation tend to be temporary: when energy and commodity prices surged, the prices group contributed at most 0.7 of the 4.9 to 5.6 percentage points by which the trends rose. This distinction matters for monetary policy, which typically looks through temporary supply shocks, although it should be clear that the groups of indicators are not structural shocks: employment, for example, reflects labour supply as well as labour demand.

Over the longer history, interest and exchange rates are the largest contributor to the fall of the trends during the disinflation of the early 1980s. The forward-looking trend started to fall in 1979, a year before inflation peaked, and rising long-term interest rates lowered it during the tightening of monetary policy led by Paul Volcker \citep{GoodfriendKing2005} and again in 2022.

We also find that the contributions are clearly nonlinear: for eight of the nine groups, a linear function of a group's own indicators cannot reproduce between 16 and 40\% of the variation of its contribution to the innovation in the encoder, and up to 53\% in the attention architecture. The nonlinearity also matters for the trend as a whole, because a linear version of the model, with a single linear layer from the indicators to the innovation, gives a trend that follows the same path but is much noisier, and whose predictive densities are less accurate out of sample.

We finally evaluate the model out of sample, during the post-pandemic surge of inflation, re-estimating it every month from December 2020 to November 2023 with the data up to that month. Given its parameters, the trend in a month is a deterministic function of the indicators up to that month, so later data revise it only through the re-estimation of the network. At the time, the model attributed the rise of the forward-looking trend mainly to employment and to money and credit, and the same groups account for most of its fall from June 2022. We then compare the forecasts of inflation of the model with those of an autoregression, of the UC-SV model of \citet{StockWatson2007}, of a random walk and of the Survey of Professional Forecasters (SPF). The forecasts of both trends are more accurate than those of UC-SV and of the random walk, one month ahead and over the next twelve months, and about as accurate as those of the autoregression, and the forward-looking trend gives the most accurate density forecasts of inflation over the next year. The forward-looking trend also signalled the rise of inflation well before the professional forecasters: in the first survey of 2021, the SPF expected inflation of 2.2\% over the next year, while the forward-looking trend forecast close to 4\%, and inflation turned out at 7.3\%. The professional forecasters, however, anticipated the fall of inflation earlier. Neither architecture is more accurate than the other at the 5\% level in this episode, so we read their different attributions of the rise of the forward-looking trend as model uncertainty.

The paper contributes to several strands of the literature. First, it contributes to the literature on trend inflation by proposing a model that bridges traditional UC models and machine learning approaches. UC models have a clear structural interpretation, but they use few variables, they are linear, and they cannot attribute the movements of the latent trend to economic drivers. Machine learning approaches are flexible, but their output often lacks a structural economic interpretation. In Deep-$\pi^*$, the innovations to the trend are a function of observables, which explains what drives the trend, and the neural network sits inside a trend-cycle decomposition, which gives its output an economic meaning. Second, it contributes to the literature on machine learning for economic and financial forecasting \citep{GouletCoulombeStevanovic2022, Medeiros2021, GooKellyXiu2020} with an architecture designed for macroeconomic time series: short samples, temporal structure and the need for economic interpretability. Unlike most of this literature, which uses neural networks as direct forecasting tools, we embed the network within a trend-cycle decomposition. Third, it contributes to the literature on attention mechanisms for time series \citep{Vaswani2017, Wen2023} with a parsimonious, similarity-based causal attention that adds three parameters, which makes it feasible for macroeconomic samples of a few hundred observations. Fourth, the two loss functions make explicit a tension in the definition of trend inflation that is usually left implicit: the trend that best explains current inflation need not coincide with the trend that best predicts future inflation. Comparing the drivers of the two trends after the pandemic shows which indicators carried forward-looking information in that episode. Fifth, it brings the volatility of inflation surprises into the same structure, with a persistent volatility state driven by a second network of indicators and past errors. As in \citet{GouletCoulombeFrenetteKlieber2026}, the volatility responds both to observed macroeconomic conditions and to past surprises; in Deep-$\pi^*$, it follows the same recursion as the trend, which gives it the persistence of stochastic volatility models while keeping the likelihood in closed form.

The remainder of the paper is organised as follows. Section~\ref{sec:literature} reviews the related literature. Section~\ref{sec:model} presents the model and the four architectures. Section~\ref{sec:estimation} discusses the estimation, the loss functions and the uncertainty bands. Section~\ref{sec:data_results} describes the data and presents the eight estimates of the trend and the volatility of inflation surprises. Section~\ref{sec:drivers} decomposes the trends into the contributions of groups of indicators: the drivers since 2019, seven episodes of US inflation history, and the nonlinearity of the contributions. Section~\ref{sec:out_of_sample} evaluates the estimates out of sample and in real time, against standard benchmarks and the forecasts of professional forecasters. Section~\ref{sec:conclusion} concludes.
 \section{Related Literature}
\label{sec:literature}

This paper relates to four literatures: econometric models of trend inflation, machine learning in macroeconomics, attention mechanisms for time series, and interpretability in machine learning. Across them, the main novelty of Deep-$\pi^*$ is to place a neural network inside a model of trend inflation, so that the innovations to the trend are a function of a large panel of macroeconomic indicators, the same model delivers a current and a forward-looking trend, and each movement of the trend can be decomposed exactly into the contributions of groups of indicators.

\subsection{Trend Inflation: Econometric Approaches}

The modern literature on trend inflation builds on the unobserved components model with stochastic volatility (UC-SV) of \citet{StockWatson2007}, which has become a workhorse in central banks. The model decomposes inflation into a trend and a serially uncorrelated transitory component, with stochastic volatility in both. Because the trend follows a random walk, it acts as a time-varying intercept, and the UC-SV model belongs to the wider family of time-varying parameter (TVP) models. TVP-VARs with a drifting intercept, such as those of \citet{Cogley2005} and \citet{primiceri2005time}, also imply a trend, the local mean of inflation, and \citet{CogleyPrimiceriSargent2010} find with such a model substantial variation over time in both trend inflation and the persistence of the US inflation gap. Later work extends this framework in several directions. \citet{ChanKoopPotter2013, ChanKoopPotter2016} bound the random walk so that the trend stays within economically reasonable values, \citet{Chan2017} lets the volatility of inflation affect its conditional mean, and \citet{ChanClarkKoop2018} link the trend to survey-based long-run inflation forecasts. Other studies sharpen the estimates of the trend with more data: \citet{StockWatson2016} use seventeen sectoral components of PCE inflation, and \citet{Mertens2016} measures of realised inflation, survey expectations and the term structure of interest rates, while \citet{ClarkDoh2014} compare alternative models of trend inflation in forecasting. In all these models, the trend is a latent state driven by its own shock. In the score-driven models of \citet{CrealKoopmanLucas2013}, by contrast, time-varying parameters are updated by the score of the likelihood, which is a function of past data, and \citet{DelleMonachePetrella2017} apply them to a time-varying autoregression for US CPI inflation in which both the coefficients and the volatility vary.

Other work relates trend inflation to economic fundamentals and to inflation expectations, often motivated by the Great Recession, when inflation fell less than many models predicted. \citet{BallMazumder2011} study the dynamics of inflation in this period through the Phillips curve, and \citet{CoibionGorodnichenko2015} explain the missing disinflation by the rise in household inflation expectations between 2009 and 2011. Over a longer period, \citet{Watson2014} documents that the persistence of US inflation has fallen markedly since the early 1980s, and \citet{Bernanke2007} reads the lower variability of trend inflation since then as evidence of better, though still imperfect, anchoring of long-run inflation expectations. In the same vein, \citet{HazellEtAl2022} estimate a small slope of the Phillips curve from US state data and attribute the greater stability of inflation since the 1990s mostly to better-anchored long-run expectations. \citet{MertensNason2020} find that longer-horizon survey forecasts inform estimates of trend inflation, whereas \citet{CecchettiFeroli2017} find that the slowly moving local mean of US inflation since 1984 responds little to labour market slack and not at all to measures of inflation expectations. Finally, \citet{Hasenzagl2022} combine the UC framework with a semi-structural Phillips curve, in which the trend is anchored by long-run inflation expectations and the cycle is linked to real activity. These models make trend inflation partly explainable, but with a small number of predetermined variables and linear relationships.

Time-varying volatility is also a standard feature of models of trend inflation. In \citet{StockWatson2007} and \citet{Cogley2005}, the log variances follow random walks. \citet{Clark2011} shows that stochastic volatility improves the real-time density forecasts of Bayesian VARs, and \citet{ChanEtAl2020} show that it also improves the density forecasts of large VARs with up to 196 variables. In the ARCH and GARCH models of \citet{Engle1982} and \citet{Bollerslev1986}, and in the exponential GARCH of \citet{Nelson1991}, the variance is instead a function of past errors, which keeps the likelihood in closed form. More recently, \citet{GouletCoulombeFrenetteKlieber2026} forecast macroeconomic volatility with a neural network that responds to observed indicators and to past forecast errors. Our volatility state combines these elements: it is persistent, as in stochastic volatility models, it is driven by observed data, as in GARCH models, and it responds to indicators and to past errors, as in \citet{GouletCoulombeFrenetteKlieber2026}.

Relative to these models, Deep-$\pi^*$ makes three changes. First, the trend moves with the state of the economy. Its innovation is a function of the indicators in the current month and, in the attention architecture, of their recent history, so the trend has no shock of its own. In this respect it is closer to the score-driven models, in which time variation is also a function of the data, than to the latent trends of UC models. Second, the function is nonlinear and estimated with neural networks, so that the data determine which indicators matter and whether their effects interact, and exact group Shapley values attribute each movement of the trend to groups of indicators. Third, the loss function states which trend is estimated. The Econometrician's loss gives the persistent component of current inflation, as in UC models, while the Central Banker's loss gives the trend that best predicts inflation over the next twelve months. This predictive definition is shared by measures of core inflation that are judged by how well they predict future headline inflation \citep{AlmuzaraSbordone2024, Wright2026, GouletCoulombeEtAl2024Core}. These measures are built from the components of the price index, whereas Deep-$\pi^*$ builds the forward-looking trend from indicators of the whole economy.

\subsection{Machine Learning in Macroeconomics}

Machine learning methods have become common in macroeconomic forecasting. \citet{Medeiros2021} show that random forests outperform standard benchmarks in forecasting US inflation. \citet{Naghi2024} replicate this result with a larger set of methods, but find that the advantage of random forests is smaller in Canadian data, in which gradient-boosted trees are more accurate, and that in US data extended to 2022 random forests forecast poorly in 2020--2022. To understand where the gains come from, \citet{GouletCoulombeStevanovic2022} decompose them into four features of machine learning: nonlinearities, regularisation, cross-validation and alternative loss functions. Nonlinearity is the main source of the gains, which are largest in periods of high macroeconomic uncertainty, financial stress and housing bubble bursts.

Neural networks have also been applied directly to inflation forecasting. \citet{AlmosovaAndresen2023} forecast monthly US CPI inflation with long short-term memory networks (LSTMs) \citep{Hochreiter1997}, a type of recurrent neural network (RNN), which slightly outperform autoregressive, Markov-switching and feed-forward network benchmarks and are about as accurate as a seasonal ARIMA model. \citet{Paranhos2023} applies LSTMs to FRED-MD data and finds that their latent factors are highly correlated with business cycle indicators, and \citet{Stoneman2024} train an RNN on 60 years of quarterly data that outperforms the Survey of Professional Forecasters and projects a spike in inflation similar to that of 2021--2022. \citet{Carriero2024} compare time series foundation models, which adapt the architectures of large language models to forecasting, with standard econometric methods on FRED-MD, and find that a hybrid, in which a foundation model forecasts the residuals of a Bayesian VAR, gives the most accurate forecasts. In all these studies, the network is the forecasting model, and its output is the forecast.

Large panels of indicators are commonly summarised by principal components \citep{StockWatson2002}, and neural networks offer nonlinear alternatives. \citet{Andreini2023} estimate deep dynamic factor models, in which neural networks extract nonlinear common factors from large macroeconomic panels, and \citet{Klieber2024} replaces the principal components of factor-augmented VARs with nonlinear factors obtained from locally linear embedding or an autoencoder. \citet{HauzenbergerHuberKlieber2023} forecast inflation in real time with factors from autoencoders and other nonlinear dimension reduction techniques, and find that they are competitive with linear approaches. The encoder of Deep-$\pi^*$ also compresses a large panel nonlinearly, into a single index. The compression is supervised, however: the index is the innovation to the trend, so the encoder retains the information in the indicators that is relevant for trend inflation.

A further line of work embeds machine learning in an economic model, so that the output of the network has an economic interpretation. \citet{GouletCoulombe2025} lets the coefficients of a macroeconomic equation vary with the state of the economy through random forests. The Hemisphere Neural Network (HNN) of \citet{GouletCoulombe2022NPC} splits its final layer into components that can be interpreted as the latent states of a neural Phillips curve, such as inflation expectations and the output gap, each fed by a group of predictors, and \citet{GouletCoulombeFrenetteKlieber2026} add a hemisphere for volatility. \citet{BuckmannPotjagailoSchnattinger2025} decompose UK inflation with boosted trees into components aligned with a hybrid Phillips curve, identify the contributions of demand and supply with monotonicity constraints, and find an L-shaped effect of labour market tightness that amplified the recent surge. Deep-$\pi^*$ takes this approach to the trend-cycle decomposition of inflation: the network sits inside the model, and its output is the innovation to persistent inflation.

\subsection{Attention Mechanisms and Transformers for Time Series}

The transformer architecture of \citet{Vaswani2017}, introduced for natural language processing, has been adapted to time series forecasting with mixed results, as the survey of \citet{Wen2023} documents. Standard attention is designed for long sequences and large data sets, and it can be too rich for macroeconomic time series. In self-attention, each output is a weighted average of learned value projections of the inputs, with weights computed from learned query and key projections. With $k$ inputs and projections that keep their dimension, the three projections alone add $3k^2$ parameters, 30,000 with a hundred indicators, while macroeconomic samples have at most several hundred monthly or quarterly observations. With so few observations for each parameter, the attention weights can overfit the sample, and the learned projections make them harder to interpret. Designs adapted to time series, such as the temporal fusion transformers of \citet{Lim2021} and PatchTST \citep{Nie2023}, still learn query and key projections and are intended for long or high-frequency series, with thousands to tens of thousands of observations in most of their applications, one to two orders of magnitude more than in macroeconomic applications.

Our attention mechanism is designed to be parsimonious and interpretable. It computes the attention weights from the cosine similarity between the indicators of different months, with a learned temperature as in the scaled cosine attention of \citet{LiuSwinV2_2022}, and a geometric decay that adds a linear recency bias to the scores, as in \citet{Press2022}, here with an estimated slope. Without learned projections, it adds only three scalar parameters to the network. The weights depend only on the indicators and on the distance between months: the decay gives more weight to recent months, and the similarity gives more weight to past episodes with similar macroeconomic conditions. Each weight therefore has a direct reading, as the share of a past month in the inputs of the current month. This reading is close to the dual interpretation of machine learning forecasts of \citet{GouletCoulombeGobelKlieber2024}, who write each prediction as a combination of past observations, with weights given by the proximity between current and past economic conditions.

\subsection{Interpretability in Machine Learning for Economics}

Interpretability is a central requirement for neural network models in economics, and two prominent approaches have emerged. \citet{Sundararajan2017} develop Integrated Gradients, an axiomatic attribution method that distributes a model's prediction among its input features by integrating gradients along a path from a baseline input. \citet{Lundberg2017} unify several attribution methods under the SHAP (SHapley Additive exPlanations) framework, which provides measures of feature importance grounded in cooperative game theory. Building on these tools, \citet{BuckmannJoseph2023} propose a workflow for machine learning models in economics based on Shapley value decompositions and Shapley regressions, and apply it to forecasting US unemployment with FRED-MD.

In macroeconomics, the HNN discussed above attributes inflation to its hemispheres, whose groups of predictors are assigned to the components in advance. In Deep-$\pi^*$, all indicators enter one network, and we measure the contribution of each group after estimation with exact Shapley values for groups of features, as in the groupShapley of \citet{JullumRedelmeierAas2021}. With nine groups, all coalitions can be evaluated, so the contributions are exact and add up in every month. We compute them for the trend innovations, and because the trend is linear in the innovations, they carry over to the \textit{level} of trend inflation. The result is an exact decomposition of the trend, month by month, into the contributions of groups of macroeconomic indicators.
 \section{Model and Architecture}
\label{sec:model}

\subsection{The Deep-\texorpdfstring{$\pi^*$}{pi*} Model}\label{subsec:model_spec}

We model observed inflation, $\pi_t$, as an autoregression whose long-run level is a persistent trend, $\pi_t^*$:
\begin{align}
    \pi_t &= \underbrace{\sum_{p=1}^{P} \phi_p \, \pi_{t-p}}_{\text{Autoregressive dynamics}} + \underbrace{\left(1 - \sum_{p=1}^{P} \phi_p\right) \pi_t^*}_{\text{Trend}} + \sigma_t \, \varepsilon_t, \quad \varepsilon_t \sim t_\nu(0,1), \label{eq:observation} \\[6pt]
    \pi_t^* &= \theta \, \pi_{t-1}^* + (1 - \theta) \, \tau_t, \label{eq:state}
\end{align}
where $\tau_t$ is the innovation to the trend and $\pi_0^*$ is an estimated initial condition. The innovation is a function of a vector of macroeconomic indicators, $\bx_t \in \R^k$, and possibly of their history:
\begin{equation}\label{eq:innovation}
    \tau_t = c \, \tanh\!\left(\frac{g(\bx_1, \bx_2, \ldots, \bx_t)}{c}\right),
\end{equation}
where $g(\cdot)$ is an unknown, possibly nonlinear function approximated by a neural network, and $c$ bounds the innovation.

Equation \eqref{eq:observation} is a standard autoregression in which the trend plays the role of a time-varying mean: in a steady state with $\pi_t = \pi_t^* = \bar{\pi}$, the equation holds for any value of the autoregressive coefficients $\{\phi_p\}$. The lags enter the equation in levels, but subtracting $\pi_t^*$ from both sides gives
\[
    \pi_t - \pi_t^* = \sum_{p=1}^{P} \phi_p \left(\pi_{t-p} - \pi_t^*\right) + \sigma_t \varepsilon_t,
\]
so the deviation of inflation from the current trend, the inflation gap, follows approximately an autoregression, since the trend moves slowly. We restrict the sum of the coefficients to the interval $(0,1)$, so that the trend enters with a positive weight and the lag polynomial has no unit root, and the estimated coefficients imply a stationary gap, so that inflation reverts to the trend. We denote the error of month $t$ by $r_t = \sigma_t \varepsilon_t$. It follows a Student-$t$ distribution with $\nu$ degrees of freedom, whose heavy tails, for a small $\nu$, limit the influence of occasional extreme months, and its scale, $\sigma_t$, is either constant, $\sigma_t = \sigma$, or a function of a time-varying volatility state with the same structure as the trend (Section~\ref{subsec:volatility}).

Equation \eqref{eq:state} is the main departure from the unobserved components literature. In that literature the trend typically follows a random walk, $\pi_t^* = \pi_{t-1}^* + \eta_t$ with $\eta_t \sim \mathcal{N}(0, \omega^2)$, so that its changes are unpredictable shocks, whereas in Deep-$\pi^*$ the innovation is a function of observable macroeconomic conditions. The parameter $\theta \in (0,1)$ controls the persistence of the trend. After $n$ months, the trend has absorbed a share $1 - \theta^n$ of a permanent shift in the innovation, so a value of $\theta$ close to one gives the slow movements of trend inflation. Solving the recursion backwards, the trend is an exponentially weighted average of past innovations:
\[
    \pi_t^* = \theta^t \pi_0^* + (1-\theta) \sum_{s=1}^{t} \theta^{t-s} \tau_s.
\]
Each innovation is therefore the level of the trend that the conditions of its month imply, and it moves the trend by $(1-\theta)(\tau_t - \pi_{t-1}^*)$ in that month. The trend, the innovations and the bound below are in the standardised units of inflation (Appendix~\ref{app:units}).

Equation \eqref{eq:innovation} adds a bound on the innovation: the hyperbolic tangent keeps it within $(-c, c)$, and for values of $g$ that are small relative to $c$, $\tau_t \approx g$, while larger values are compressed towards the bound. The bound therefore limits how fast the trend can move in a month:
\begin{equation}\label{eq:speed}
    |\pi_t^* - \pi_{t-1}^*| = (1-\theta)\, \big|\tau_t - \pi_{t-1}^*\big| \le (1-\theta)\big(c + |\pi_{t-1}^*|\big).
\end{equation}
We set $c$ above the usual size of the innovations, which move much more than the trend that averages them, so that the bound compresses most innovations little and caps the monthly move of the trend at a value that is large for trend inflation (Section~\ref{subsec:settings}). The bound is most useful at the end of an estimation sample. In the middle of the sample, a large move of the trend must also fit the months that follow, which pull it back if the move was transitory. At the end of the sample, however, those months are not yet observed, so a few extreme months could move the trend far, as at the end of 2008, when monthly inflation was between $-10\%$ and $-21\%$ annualised from October to December. By capping the monthly move, the bound keeps the trend a slowly moving component in real time as well. The bound is in the spirit of the bounded trend inflation of \citet{ChanKoopPotter2013, ChanKoopPotter2016}, who bound the level of the trend. 

\paragraph{Connection to recurrent neural networks.} Taken together, equations \eqref{eq:state} and \eqref{eq:innovation} can be viewed as a restricted recurrent neural network (RNN). In a general RNN, a hidden state evolves as $s_t = f(s_{t-1}, \bx_t; \bm{W})$ for some function $f$ approximated by a neural network. In our specification, the hidden state is $\pi_t^*$, and the recurrence is a convex combination of the previous state and a bounded function of the inputs, as in equations \eqref{eq:state} and \eqref{eq:innovation}. This structure delivers the properties of a trend: persistence, through $\theta$ close to one, and convergence to a level determined by the data, through the term $(1-\theta)\tau_t$. It also guarantees that the hidden state can be read as trend inflation.

It remains to specify the function $g(\cdot)$ and the scale of the errors, $\sigma_t$. We consider four architectures, which combine two choices. The first choice is how the indicators enter the trend: either through an encoder that maps the indicators of each month to the innovation (Section~\ref{subsec:encoder}), or through an attention layer that first forms a weighted average of the indicators of the current and past months, followed by the same encoder (Section~\ref{subsec:attention}). The second choice is the scale of the errors, which is either constant or follows a volatility state driven by a second neural network of the same indicators and of past errors (Section~\ref{subsec:volatility}).

\subsection{Architecture I: The Encoder}\label{subsec:encoder}

\begin{figure}[H]
    \centering
    \includegraphics{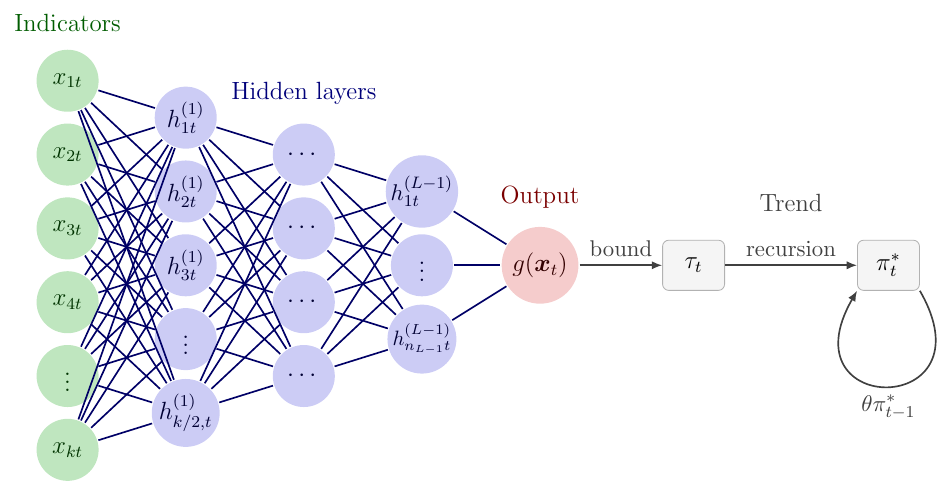}
    \caption{\textbf{The encoder.} The $k$ indicators of month $t$ pass through $L-1$ hidden layers with SiLU activations, whose widths halve from one layer to the next, to a single linear output, $g(\bx_t)$. The bound in equation~\eqref{eq:innovation} maps the output to the innovation, $\tau_t$. The recursion in equation~\eqref{eq:state} combines the innovation with the trend of the previous month, $\theta \pi^*_{t-1}$, to give the trend, $\pi^*_t$.}
    \label{fig:encoder}
\end{figure}

The encoder is a feedforward neural network, or multi-layer perceptron \citep[Chapter 6]{Goodfellow2016}, whose layer widths decrease progressively. It maps the indicators of month $t$ to the network output:
\begin{equation}\label{eq:encoder}
    g(\bx_t) = g_L \circ g_{L-1} \circ \cdots \circ g_1(\bx_t),
\end{equation}
where each hidden layer $g_\ell(\cdot) = a(W_\ell \cdot + b_\ell)$ applies a linear transformation followed by a nonlinear activation function $a(\cdot)$, and where the last layer is linear. As the activation, we use the Sigmoid Linear Unit, $\text{SiLU}(x) = x/(1 + e^{-x})$ \citep{HendrycksGimpel2016, Elfwing2018}, also known as Swish \citep{Ramachandran2018}, a smooth function with well-behaved gradients. The width halves from one layer to the next. This progressive compression can be read as a nonlinear analogue of principal component analysis. The first layer forms about $k/2$ nonlinear combinations of the $k$ indicators, the second condenses them into about $k/4$, and so on, until a single output remains. Figure~\ref{fig:encoder} shows the architecture graphically.

\subsection{Architecture II: The Attention Mechanism}\label{subsec:attention}

A limitation of the encoder is that it reads only the indicators of the current month (Figure~\ref{fig:encoder}). Some indicators, however, may affect persistent inflation with a delay. Money is the classic example: \citet{Friedman1961} argues that monetary actions affect economic conditions only after long and variable lags, so the surge in the monetary aggregates in the spring of 2020, for instance, could matter for the trend well after it happened. Adding lags of the indicators to the encoder would require fixing their number in advance, and it would multiply the number of inputs and parameters. We therefore use a simplified attention mechanism, which forms a weighted average of the indicators of the current and past months without fixing the number of lags. Its weights favour the past months whose macroeconomic conditions resemble those of the current month, and they give more weight to the recent past.  

Attention mechanisms are the basis of modern language models, in which the meaning of a word depends on the words around it \citep{Vaswani2017}. For each word, attention builds the input of the network as a weighted average of the words of the text, with weights that depend on how relevant each of them is to the current one, so that the network can draw on the parts of the text that matter most. In a time series, the words are the months. For the trend of a given month, our attention looks back over the past months, compares the economic conditions of each of them with those of the current month, and gives more weight to the months that look alike and to the more recent ones. The encoder then reads this weighted summary of the current and past months.

\subsubsection{Similarity-Based Causal Attention}

For month $t$ and a month $i$, we construct the attention weight as
\begin{equation}\label{eq:attention_weights}
    \alpha_{t,i} = \begin{cases}
        \dfrac{\tilde{\alpha}_{t,i}}{\sum_{j \leq t} \tilde{\alpha}_{t,j}} & \text{if } i \leq t, \\[10pt]
        0 & \text{if } i > t,
    \end{cases}
\end{equation}
where the unnormalised weights combine a similarity score and a geometric decay:
\begin{equation}\label{eq:unnormalized_weights}
    \tilde{\alpha}_{t,i} = \underbrace{\exp\big(\beta \cos(\bx_t, \bx_i)\big)}_{\text{similarity}} \times \underbrace{\delta^{\,t-i}}_{\text{decay}}, \qquad \cos(\bx_t, \bx_i) = \frac{\bx_t' \bx_i}{\lVert \bx_t \rVert \, \lVert \bx_i \rVert}.
\end{equation}
The sharpness $\beta > 0$ and the decay $\delta \in (0,1)$ are estimated. Three features of the mechanism matter for our application: it compares the patterns of the indicators, its memory decays with time, and it only looks back. We discuss them in turn.

\paragraph{Similarity of patterns.} The weight of a past month grows with the similarity between its indicators and those of the current month, because a past month is more informative about the current trend when the economy was then in a similar state. The effect of strong money growth on the trend, for example, may differ between a tight and a slack labour market, and weighting the past months by their similarity lets the trend draw mainly on the months that share the current combination of conditions. These months need not be recent: with a decay close to one, the trend in 2021 could draw on an episode such as the mid-1970s if the indicators of the two periods were similar. How far back the attention looks in practice therefore depends on the estimated decay, which Section~\ref{subsec:trends} reports.

We measure the similarity with the cosine of the angle between the indicators of the two months, which compares their pattern independently of their overall size. Each month thus has a similarity of one with itself and with any past month that shows the same pattern, so that a month with extreme values, such as November 2008 or April 2020, can still draw on the past months whose indicators move in the same pattern. The sharpness $\beta$ then controls how strongly the weights concentrate on the most similar months.

\paragraph{Geometric temporal decay.} The factor $\delta^{t-i}$ gives the attention a memory that decays exponentially. For $\delta = 0.9$, the weight of a month one year in the past is multiplied by $0.9^{12} \approx 0.28$, and that of a month five years in the past by $0.9^{60} \approx 0.002$. Because the decay multiplies the similarity score, however, a distant but highly similar episode can still receive substantial weight.

\paragraph{Strict timing causality.} Setting $\alpha_{t,i} = 0$ for $i > t$ makes the innovation in month $t$ depend only on information available up to month $t$. The mask is needed because the trend must be computable in real time, from the data available at each date, so the attention looks back but never forward.

\paragraph{Relation to transformer attention.} Finally, our attention mechanism is the scaled dot-product attention of \citet{Vaswani2017} with three changes. First, the query, key and value maps are the identity. This attention mechanism therefore adds only three scalar parameters ($\beta$, $\delta$ and the output scale $\gamma$ below), while learned projections would add tens of thousands. Second, the scores use the cosine similarity with a learned temperature, as in the scaled cosine attention of \citet{LiuSwinV2_2022}. Third, the decay adds a linear recency bias, $(t-i)\log\delta$, to the scores, as in the attention with linear biases of \citet{Press2022}, here with an estimated slope.

\subsubsection{The Attend-First Architecture}

\begin{figure}[H]
    \centering
    \includegraphics[width=\textwidth]{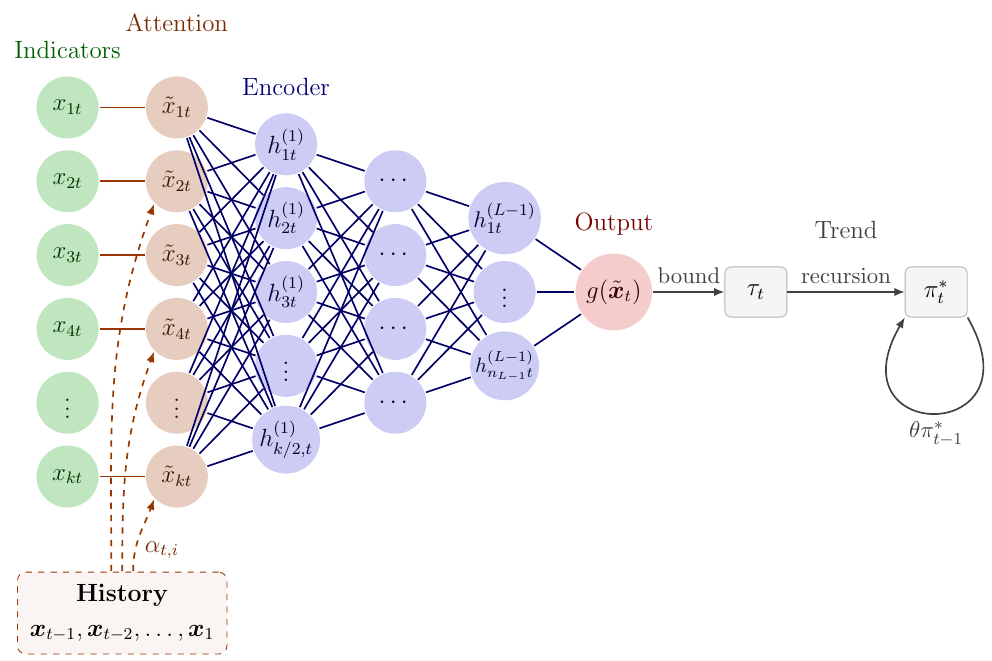}
    \caption{\textbf{The attention architecture.} The attention layer replaces each indicator by a weighted average of its values in the current and past months, $\tilde{\bx}_t = \gamma \sum_{i \le t} \alpha_{t,i} \bx_i$. The weight $\alpha_{t,i}$ depends on the similarity between the indicators of months $t$ and $i$ and on the number of months between them, as in equation~\eqref{eq:unnormalized_weights}. The attention layer has no weight matrices and adds three parameters: $\beta$, $\delta$ and $\gamma$. The encoder of Figure~\ref{fig:encoder} then maps $\tilde{\bx}_t$ to the output, $g^{\text{enc}}(\tilde{\bx}_t)$, and the bound and the recursion give the innovation, $\tau_t$, and the trend, $\pi^*_t$, as in that figure.}
    \label{fig:attention}
\end{figure}

For each month $t$, the attention layer thus produces a weighted average of the indicators of the current and past months:
\begin{equation}\label{eq:attention_output}
    \tilde{\bx}_t = \gamma \sum_{i=1}^{t} \alpha_{t,i} \, \bx_i,
\end{equation}
where $\gamma$ is an estimated output scale. The encoder of Section~\ref{subsec:encoder} then maps $\tilde{\bx}_t$ to the network output, so that $g(\bx_1, \ldots, \bx_t) = g^{\text{enc}}(\tilde{\bx}_t)$. The encoder has the same layers in the two architectures, which therefore differ only by the attention layer. Figure~\ref{fig:attention} shows the full architecture with the attention layer followed by the encoder.

\subsection{Architecture for the Volatility of Inflation Surprises}\label{subsec:volatility}

The size of inflation surprises, the errors $r_t$ of equation~\eqref{eq:observation}, may also change over time. Two of our four architectures therefore let the scale of the errors vary, following a volatility state that is built in the same way as the trend and whose innovations come from a second network.

\subsubsection{The Volatility State}

The scale of the errors in equation~\eqref{eq:observation} is
\begin{equation}\label{eq:scale}
    \sigma_t = \bar{\sigma} \exp(v_t), \qquad v_t = \bar{v} \tanh\!\left(\frac{\lambda_t}{\bar{v}}\right),
\end{equation}
where $\bar{\sigma}$ is an estimated baseline scale, $\lambda_t$ is a volatility state and $\bar{v} > 0$ bounds the log scale. The state is as persistent as the trend:
\begin{equation}\label{eq:volstate}
    \lambda_t = \theta \, \lambda_{t-1} + \bar{\xi} \, \xi_t, \qquad \xi_t = \tanh\!\big(q(\bm{z}_t)\big), \qquad \lambda_0 = 0,
\end{equation}
with the same persistence, $\theta$, and a fixed bound $\bar{\xi} > 0$ on its monthly change. The innovation $\xi_t \in (-1, 1)$ comes from a second, small neural network $q(\cdot)$ of the inputs $\bm{z}_t$ described below. This structure has three properties.

First, there is a limit on the speed of volatility. Because $|\xi_t| < 1$, the state moves in a month by less than $\bar{\xi}$ from its decayed value, $|\lambda_t - \theta \lambda_{t-1}| < \bar{\xi}$, so with a small value of $\bar{\xi}$ the scale can only move slowly. The volatility state thus follows the slow movements in the size of inflation surprises, in the same way as the trend follows the slow movements in the level of inflation.

Second, the volatility process is bounded. The hyperbolic tangent in equation~\eqref{eq:scale} keeps $\sigma_t$ between $\bar{\sigma} e^{-\bar{v}}$ and $\bar{\sigma} e^{\bar{v}}$, whatever the value of the state. For small values of the state, $v_t \approx \lambda_t$.

Finally, the state is as persistent as the trend, with a half-life of $\log 0.5 / \log \theta$ months, which is long for $\theta$ close to one. When we forecast from month $t$, the inputs of the later months are not available, and we set their innovations to zero, so that the state decays geometrically from its current value, as the trend does: its forecast $h$ months ahead is $\theta^h \lambda_t$.

\subsubsection{Inputs and Timing}

The inputs to the volatility network combine the macroeconomic indicators that also enter the trend network and the absolute values of the last $P_\sigma$ errors of the inflation equation, divided by $\bar{\sigma}$. The indicators let the volatility move with observed macroeconomic conditions before large surprises arrive, while the past errors let it respond to the surprises themselves; \citet{GouletCoulombeFrenetteKlieber2026} call these two sources proactive and reactive volatility. 

Through the past errors, the volatility state has the structure of the ARCH and GARCH models \citep{Engle1982, Bollerslev1986}, in which the variance of the errors is a function of past residuals. As in the exponential GARCH of \citet{Nelson1991}, the state drives the log of the scale. The state is also the observation-driven counterpart of the stochastic volatility with random-walk log variances that is standard in models of trend inflation \citep{Cogley2005, primiceri2005time, StockWatson2007}. In those models, the log variance has its own unobserved shock, and estimation requires filtering or simulation. In Deep-$\pi^*$, the innovation $\xi_t$ is a function of observed data, so the likelihood is available in closed form, and the volatility network is estimated jointly with the trend by gradient descent. With $\theta$ close to one, the state is close to a random walk, and the bound in equation~\eqref{eq:scale} keeps the scale within a fixed range.

\subsubsection{The Volatility Network}

The network $q(\cdot)$ is a small feedforward network with $k + P_\sigma$ inputs, a few narrow hidden layers with SiLU activations, and a single output. Like the trend network, it is regularised with dropout (Section~\ref{subsec:regularization}). 

\begin{figure}[H]
    \centering
    \includegraphics{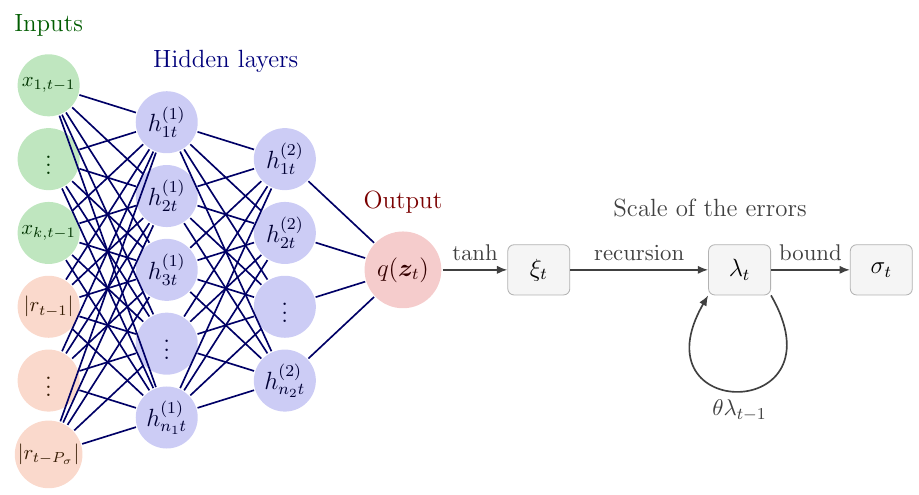}
    \caption{\textbf{The volatility network.} The inputs are the indicators of month $t-1$ (green) and the absolute values of the last $P_\sigma$ errors, divided by $\bar{\sigma}$ (orange). Hidden layers with SiLU activations map them to the output, $q(\bm{z}_t)$. The innovation $\xi_t = \tanh(q(\bm{z}_t))$ enters the volatility state $\lambda_t$ through a recursion with the same persistence as the trend, equation~\eqref{eq:volstate}, and the bound in equation~\eqref{eq:scale} maps the state to the scale, $\sigma_t$. The inputs shown are those of the Econometrician's loss (Section~\ref{subsec:loss}). Under the Central Banker's loss, they are the indicators of month $t$ and the one-month-ahead forecast errors up to month $t$.}
    \label{fig:volatility}
\end{figure}

Figure~\ref{fig:volatility} shows the network and the steps from its output to the scale of the errors.

 \section{Estimation}
\label{sec:estimation}

\subsection{Parameters to Estimate}

The model contains the following groups of parameters, collectively denoted $\Theta$:
\begin{enumerate}[label=(\roman*)]
    \item \textit{Neural network weights and biases}: all parameters of the encoder and, in the attention architecture, the sharpness $\beta$, the decay $\delta$ and the output scale $\gamma$;
    \item \textit{Autoregressive coefficients}: $\{\phi_1, \ldots, \phi_P\}$, which govern the dynamics of the gap;
    \item \textit{Scale parameters}: $\sigma > 0$ in the architectures with constant volatility. In those with time-varying volatility, the baseline scale $\bar{\sigma} > 0$, the weights and biases of the volatility network $q(\cdot)$ and, under the Central Banker's loss, the horizon factors $\kappa_2, \ldots, \kappa_H$;
    \item \textit{Initial condition}: $\pi_0^*$, the starting value of the trend. The volatility state starts at $\lambda_0 = 0$ (equation~\eqref{eq:volstate}), so the scale in the first month is within about 10\% of the estimated baseline, $\bar{\sigma}$, which plays the role of the initial condition of the volatility.
\end{enumerate}
The persistence $\theta$, the bound $c$, the bound $\bar{\xi}$ on the monthly change of the volatility state, the range $\bar{v}$ of the log scale and the degrees of freedom $\nu$ are fixed; Section~\ref{subsec:settings} gives their values in our application. Calibrating $\theta$ identifies the decomposition of inflation into trend and gap.

\subsection{Loss Functions}\label{subsec:loss}

We estimate the parameters of the model by minimising one of two loss functions. Each loss defines a different concept of trend inflation, and since both concepts matter for policy, we report the estimates under both.

\subsubsection{The Econometrician's Loss: the Current Trend}

The first loss is the negative log-likelihood of the observation equation:
\begin{equation}\label{eq:loss_baseline}
    \mathcal{L}_0(\Theta) = -\frac{1}{T}\sum_{t=1}^{T} \log \, p\!\left(r_t \,\big|\, \Theta, \nu\right),
\end{equation}
where $r_t = \pi_t - \sum_{p=1}^{P} \phi_p \pi_{t-p} - \left(1 - \sum_{p=1}^{P} \phi_p\right) \pi_t^*$ is the error of equation~\eqref{eq:observation} and $p(\cdot \,|\, \Theta, \nu)$ is the density of a Student-$t$ distribution with $\nu$ degrees of freedom and scale $\sigma_t$, constant or given by equation~\eqref{eq:scale}. This loss fits current inflation, given the indicators up to the current month. We call the resulting trend the \textit{current trend}, $\pi_t^{*,\mathrm{cur}}$: the persistent component of inflation today.

\subsubsection{The Central Banker's Loss: the Forward-Looking Trend}

The second loss evaluates the model's forecasts of inflation over a horizon of $H$ months:
\begin{equation}\label{eq:loss_forecast}
    \mathcal{L}_H(\Theta) = -\frac{1}{N}\sum_{t=1}^{T-1} \; \sum_{h=1}^{\min(H,\, T-t)} \log \, p\!\left(r_{t,t+h} \,\big|\, \Theta, \nu\right),
\end{equation}
where $r_{t,t+h} = \pi_{t+h} - \hat{\pi}_{t,t+h}$ is the $h$-step-ahead forecast error, $\hat{\pi}_{t,t+h}$ is the model's iterated forecast of inflation in month $t+h$ made in month $t$, and the density has scale $\sigma$, or $\sigma_{t,h}$ from equation~\eqref{eq:scale_h} below in the architectures with time-varying volatility. The indicators of the months after $t$ are not available, and setting their innovations to zero, the median of inflation in the standardised units (Appendix~\ref{app:units}), gives the trend forecast $\hat{\pi}^*_{t,t+h} = \theta^h \, \pi_t^*$, and the inflation forecast combines it with the autoregressive dynamics:
\[
\hat{\pi}_{t,t+h} = \left(1 - \textstyle\sum_p \phi_p\right) \theta^h \pi_t^* + \sum_{p=1}^P \phi_p \, \hat{\pi}_{t,t+h-p},
\]
where the lags start from the observed inflation of month $t$ and of the months before it. The sum in equation~\eqref{eq:loss_forecast} runs over every forecast whose outcome is in the sample, where $N = (T-H)H + H(H-1)/2$. Every month up to $T-H$ enters with its $H$ horizons, and each of the last $H-1$ months with the horizons that are already observed, which we call partial horizons. In the full sample these last months account for less than 1\% of the forecasts, but at the end of a sample available in real time they let the network learn from what followed the most recent indicators (Section~\ref{sec:out_of_sample}).

This loss rewards a trend that predicts inflation over the next $H$ months. We call the resulting trend the \textit{forward-looking trend}, $\pi_t^{*,\mathrm{fwd}}$: the trend inflation today that gives the most accurate forecasts of inflation over the next $H$ months, made with the information up to today. It is the concept most directly relevant for a policymaker who sets policy for the medium term.

\begin{figure}[H]
    \centering
    \begin{subfigure}[b]{0.48\textwidth}
        \centering
        \includegraphics[page=1]{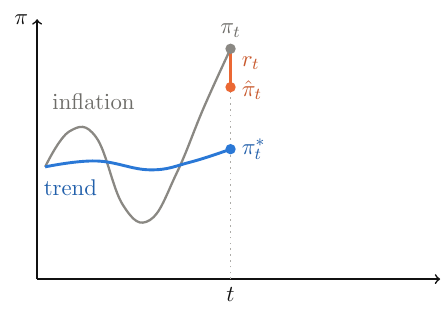}
        \caption{Econometrician's loss}
    \end{subfigure}
    \hfill
    \begin{subfigure}[b]{0.48\textwidth}
        \centering
        \includegraphics[page=2]{pictures/diagrams/losses.pdf}
        \caption{Central Banker's loss}
    \end{subfigure}
    \caption{\textbf{The two loss functions.} Inflation (grey) and the trend (blue) up to month $t$. Panel (a): the residual $r_t$ (orange) is the difference between inflation in month $t$ and its fit, $\hat{\pi}_t = \sum_p \phi_p \pi_{t-p} + (1 - \sum_p \phi_p)\pi_t^*$, from equation~\eqref{eq:observation}. Panel (b): the forecasts made in month $t$, $\hat{\pi}_{t,t+h}$ (dashed orange), and their errors, $r_{t,t+h}$, against the inflation of months $t+1, \ldots, t+H$ (grey dots). The dashed blue line is the forecast of the trend, $\theta^h \pi_t^*$.}
    \label{fig:losses}
\end{figure}

Figure~\ref{fig:losses} illustrates the two losses. The Econometrician's loss compares inflation in month $t$ with its fit, which combines the lags of inflation and the trend. The Central Banker's loss compares the inflation of the next $H$ months with the forecasts made in month $t$. As the horizon grows, these forecasts move from current inflation towards the forecast of the trend, $\theta^h \pi_t^*$.

\paragraph{The volatility state under the two losses.} In the architectures with time-varying volatility (Section~\ref{subsec:volatility}), the scale of each error uses only information available before that error is observed. The two losses evaluate different errors, so the timing of the volatility state differs:
\begin{itemize}
    \item Under the Econometrician's loss, $\sigma_t$ is the scale of the error of month $t$. Its state uses the indicators of month $t-1$ and the errors up to month $t-1$: $\bm{z}_t = \big(\bx_{t-1}, \, |r_{t-1}|/\bar{\sigma}, \ldots, |r_{t-P_\sigma}|/\bar{\sigma}\big)$, where $r_t$ is the residual of equation~\eqref{eq:loss_baseline}.\footnote{We use the indicators of month $t-1$ so that the scale of month $t$ is a forecast of the volatility of that month, made before its error is observed, as in GARCH and stochastic volatility models and as under the Central Banker's loss. With the indicators of month $t$, the scale could adapt to the month it covers, and the volatility network could lower the weight of individual months in the estimation of the trend.}
    \item Under the Central Banker's loss, the forecasts made in month $t$ use the state $\lambda_t$, which uses the indicators of month $t$ and the one-month-ahead forecast errors up to month $t$. The scale of the forecast error $h$ months ahead is
    \begin{equation}\label{eq:scale_h}
        \sigma_{t,h} = \bar{\sigma} \exp\!\Big(\bar{v} \tanh\!\big(\theta^h \lambda_t / \bar{v}\big) + \kappa_h\Big), \qquad \kappa_1 = 0.
    \end{equation}
    The formula has two parts. The first is the forecast of the volatility state: with the innovations of the months after $t$ set to zero, the state decays geometrically from its current value, $\lambda_t$, to $\theta^h \lambda_t$ after $h$ months, as the trend does (Section~\ref{subsec:volatility}). The second is a horizon factor, $\kappa_h$, which scales the error of a forecast $h$ months ahead relative to the one-month-ahead error. The scale can change with the horizon because a forecast of inflation in month $t+h$ is exposed to the surprises of all the months in between, which the autoregressive dynamics carry forward. The factors $\kappa_2, \ldots, \kappa_H$ are estimated, and $\kappa_1 = 0$ makes the scale one month ahead the one that the state implies.
\end{itemize}

\subsection{Regularisation}\label{subsec:regularization}

The networks have many parameters relative to the number of months in a macroeconomic sample. Without regularisation, they could fit the noise of individual months, and the trend would lose the persistence that defines it. We therefore use three forms of regularisation:
\begin{enumerate}[label=(\roman*)]
    \item \textit{Dropout} after each hidden layer of the encoder and of the volatility network. In the attention architecture, dropout also applies to the attention weights, with a new mask in each month.
    \item \textit{Weight decay} on the weight matrices of the networks, including the volatility network. The biases, the autoregressive coefficients, $\sigma$ or $\bar{\sigma}$, the horizon factors $\kappa_h$, $\pi_0^*$ and the attention parameters $\beta$, $\delta$ and $\gamma$ are not decayed.
    \item \textit{A prior on the sum of the autoregressive coefficients}, $\sum_p \phi_p \sim \mathcal{N}(\mu_\phi, s_\phi^2)$, which centres the persistence of the gap at a moderate value. The prior enters each objective with the weight of one month: $1/T$ under the Econometrician's loss, and $H/N$ under the Central Banker's loss, in which each month contributes up to $H$ forecasts.
\end{enumerate}

\subsection{Training}\label{subsec:training}

We estimate the parameters by gradient descent, as is standard for neural networks. In each step, the network computes the innovations of all months, the recursions in equations~\eqref{eq:state} and~\eqref{eq:volstate} turn them into the trend and the volatility, and the loss compares the implied fit, or the implied forecasts, with inflation; backpropagation through the network and the recursion then gives the gradient of the loss with respect to all the parameters. Because the trend in a month depends on the innovations of all earlier months, every step uses the whole sample. We use the AdamW optimiser \citep{Loshchilov2019} with a learning rate that is constant at first and then decays to a small value. Appendix~\ref{app:training} gives the details of the procedure.

\subsection{Uncertainty Bands and Point Estimates}\label{subsec:uncertainty}

We measure the uncertainty around the trend with Monte Carlo (MC) dropout \citep{GalGhahramani2016}. If dropout stays active after training, each pass of the data through the network switches off a different random subset of units, and \citet{GalGhahramani2016} show that the passes can then be read as draws from an approximate posterior distribution of the network. Each pass gives one path of the trend, and the 80\% band runs from the 10th to the 90th percentile of the paths. 

The point estimate of the trend comes from the network with dropout switched off. The decompositions of Section~\ref{sec:drivers} use the same network, so they add up exactly to the point estimate, which lies within its 80\% band in every month of every estimate. Finally, each estimate averages several networks trained from different random starting values, a deep ensemble \citep{Lakshminarayanan2017}. Its trend is the average of the point estimates of its networks, and its decomposition, the average of the decompositions of its networks, adds up exactly to its trend.
 \section{Data and Estimates of Trend Inflation}
\label{sec:data_results}

\subsection{Data}\label{subsec:data}

The indicators $\bx_t$ that we use to estimate the model come from the FRED-MD database \citep{McCrackenNg2016}, a monthly panel of US macroeconomic time series maintained by the Federal Reserve Bank of St.\ Louis. We use the vintage of September 2026. The panel spans eight broad categories: (i) output and income (e.g., industrial production, real personal income); (ii) labour market (e.g., unemployment rate, payroll employment, average hours); (iii) housing (e.g., housing starts, building permits); (iv) consumption, orders, and inventories; (v) money and credit (e.g., M1, M2, consumer credit, monetary base); (vi) interest rates and exchange rates (e.g., Federal funds rate, Treasury yields, term spreads); (vii) prices (e.g., PPI commodities, CPI sub-indices); and (viii) stock market (e.g., S\&P 500, dividend-price ratio).

All series are transformed to stationarity following the recommendations of \citet{McCrackenNg2016}. The target, CPI inflation, is the monthly log change of the CPI, and it is excluded from the indicators. 

Thirteen consumer price series refer to the same month as the CPI and are published with it or later: components of the CPI, the CPI excluding food, shelter or medical care, and the PCE price indexes. They therefore enter the indicators with a one-month lag, so that the trend in month $t$ is built from the macroeconomic conditions of the month and from the prices of earlier months.

After these steps, the panel has $k = 122$ indicators,\footnote{The observations for October 2025 are missing in 21 series, because the US government shutdown suspended the collection of the CPI and of the household survey. We fill one-month gaps inside series by interpolation before the transformations: geometrically for series in logarithms and linearly otherwise, which splits the change over the two months evenly. The same rule fills April 2020 in two commercial paper series. Three series that start after the beginning of the sample are not used.} and the estimation sample runs from February 1970 to December 2025, which provides $T = 671$ monthly observations. This is a moderately sized sample by the standards of macroeconomic time series but small relative to the typical training sets in the deep learning literature, which underscores the importance of the regularisation and the parsimonious architecture described above.

\subsection{Settings of the Application}\label{subsec:settings}

This section describes the main settings of our application, mostly related to the size of the networks and the calibrated parameters. Table~\ref{tab:hyperparameters} in the Appendix lists all the settings.

\paragraph{Networks.} All the networks read the $k = 122$ indicators. The encoder has five hidden layers whose widths halve from one layer to the next, $122 \to 61 \to 30 \to 15 \to 7 \to 3 \to 1$, and the attention architecture uses the same layers after its attention layer. The volatility network is smaller, with two hidden layers, $(k + P_\sigma) \to 8 \to 4 \to 1$, and it also reads the absolute values of the last $P_\sigma = 12$ errors. The inflation equation has $P = 12$ lags, which capture the dynamics of the gap at the monthly frequency, and the Central Banker's loss uses a horizon of $H = 12$ months. 

\paragraph{Calibrated parameters.} We calibrate $\theta = 0.995$, so that the trend absorbs about 63\% of a permanent shift in the innovation within 200 months, or roughly 17 years, in line with the slow movements of trend inflation documented in the literature. We set the bound at $c = 40$ standardised units, about 155 percentage points of annualised inflation. This value lies above the usual size of the innovations, which move much more than the trend that averages them, so the bound compresses most innovations little, and it limits the monthly move of the trend to about 0.8 percentage points of annualised inflation. The volatility state uses the same $\theta$, which gives it a half-life of about 138 months and a forecast that changes little over a year ($0.995^{12} \approx 0.94$). We set $\bar{\xi} = 0.1$, so that the state moves by less than 0.1 in a month from its decayed value and the scale by at most about 10\% a month, and $\bar{v} = \log 5$, so that the scale stays between $\bar{\sigma}/5$ and $5\bar{\sigma}$. Finally, the errors follow a Student-$t$ distribution with $\nu \approx 3.05$ degrees of freedom, whose heavy tails limit the influence of extreme months.

\paragraph{Estimation.} Table~\ref{tab:hyperparameters} gives the dropout rates, the weight decay, the prior on the autoregressive coefficients and the settings of the optimiser. On one CPU thread, training one network takes about 5 to 15 minutes for the encoder and 17 to 27 minutes for the attention architecture.

We estimate each of the four architectures of Section~\ref{sec:model} with both loss functions, which gives eight estimates of the trend. Each estimate is the average of five networks trained from different random starting values, a deep ensemble \citep{Lakshminarayanan2017}, which reduces the variation of the estimates that comes from the random initialisation of the weights.

The figures in this section and the next show the two architectures with time-varying volatility, and the Appendix shows those with constant volatility. The volatility changes the trends little (Figure~\ref{fig:trends_volatility} in the Appendix). Its role is in the predictive densities, which we assess in the out-of-sample evaluation of Section~\ref{sec:out_of_sample}.

\subsection{The Current and the Forward-Looking Trends}\label{subsec:trends}

\begin{figure}[htbp]
    \centering
    \includegraphics[width=\textwidth]{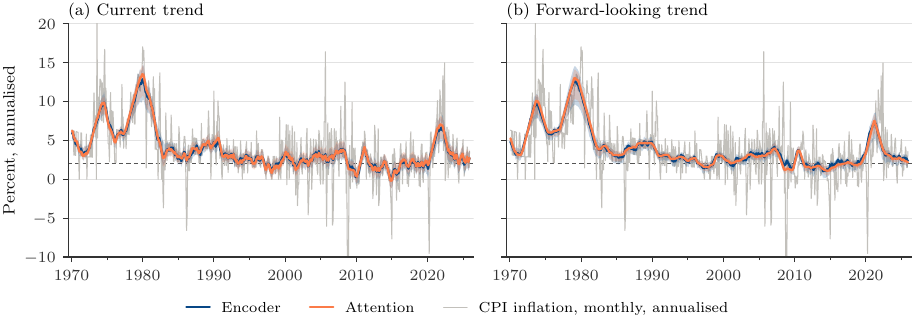}
    \caption{\textbf{The current and the forward-looking trends.} Deep-$\pi^*$ of the encoder (blue) and of the attention architecture (orange), both with time-varying volatility, with their 80\% MC-dropout bands: (a) the Econometrician's loss; (b) the Central Banker's loss. Monthly CPI inflation, annualised, in grey; the dashed line marks 2\%. Percent, annualised. The vertical axis cuts off three months of inflation: August 1973, and October and November 2008.}
    \label{fig:trends}
\end{figure}

Figure~\ref{fig:trends} shows the current trend, from the Econometrician's loss, and the forward-looking trend, from the Central Banker's loss, for the encoder and the attention architecture, together with monthly CPI inflation. Table~\ref{tab:trend_summary} summarises the four trends.

\begin{table}[htbp]
    \centering
    \small
    \caption{The trends in numbers}
    \label{tab:trend_summary}
    \begin{tabular}{lrrrr}
\toprule
 & \multicolumn{2}{c}{Current trend} & \multicolumn{2}{c}{Forward-looking trend} \\
\cmidrule(lr){2-3} \cmidrule(lr){4-5}
 & Encoder & Attention & Encoder & Attention \\
\midrule
\multicolumn{5}{l}{\textit{Levels}} \\
Peak, 1978--1981 & 12.9 & 13.5 & 12.6 & 13.0 \\
Month of the peak, 1978--1981 & Feb 1980 & Jan 1980 & Mar 1979 & Feb 1979 \\
Average, 1985--2007 & 2.9 & 2.9 & 3.0 & 3.0 \\
Lowest, 2008--2010 & 0.3 & 0.3 & 1.4 & 1.2 \\
Month of the lowest value & Dec 2009 & Feb 2010 & Oct 2009 & Sep 2009 \\
Average, 2010--2019 & 1.7 & 1.8 & 2.1 & 1.9 \\
Peak, 2020--2023 & 6.7 & 7.1 & 7.2 & 7.5 \\
Month of the peak, 2020--2023 & Mar 2022 & Nov 2021 & Mar 2021 & Mar 2021 \\
December 2025 & 2.7 & 2.7 & 2.4 & 2.2 \\
\midrule
\multicolumn{5}{l}{\textit{Smoothness and differences}} \\
Std.\ dev.\ of monthly changes & $0.28$ & $0.23$ & $0.18$ & $0.14$ \\
Width of the 80\% band & $1.93$ & $1.83$ & $1.56$ & $1.16$ \\
Mean abs.\ difference, architectures & $0.17$ & $0.17$ & $0.18$ & $0.18$ \\
Mean abs.\ difference, losses & $0.78$ & $0.74$ & $0.78$ & $0.74$ \\
Correlation of levels, losses & $0.92$ & $0.93$ & $0.92$ & $0.93$ \\
Correlation of changes, losses & $0.10$ & $0.35$ & $0.10$ & $0.35$ \\
\bottomrule
\end{tabular}
 
    \medskip
    \begin{minipage}{0.95\textwidth}
        \footnotesize \textit{Note:} Deep-$\pi^*$ with time-varying volatility. Levels in percent, annualised. Standard deviation of the monthly changes of the trend, average width of its 80\% band, and mean absolute difference from the trend of the other architecture (same loss) and of the other loss (same architecture), in percentage points. Correlations of the levels and of the monthly changes with the trend of the other loss, February 1970 to December 2025.
    \end{minipage}
\end{table}

The trend rises through the 1970s and peaks at 12.6\% to 13.5\% under both losses: in early 1980 for the current trend and in early 1979 for the forward-looking trend. The UC-SV model of \citet{StockWatson2007}, estimated on the same data, gives a trend that peaks at the same time, in January 1980, but higher, at 14.5\% (Figure~\ref{fig:trends_ucsv} in the Appendix). The trend falls steadily during the disinflation of the early 1980s, when the Federal Reserve under Paul Volcker tightened monetary policy sharply. From 1985 to 2007, during the Great Moderation, it averages about 3\% and stays mostly between 2\% and 4\%. This stability is consistent with the view that long-run inflation expectations became much better anchored after the early 1980s \citep{Bernanke2007}. 

The trend falls during the financial crisis, when economic activity declined sharply, unemployment doubled and 12-month CPI inflation fell from above 5\% in the summer of 2008 to about $-2$\% a year later. Its lowest values, between 0.3\% and 1.4\%, come between September 2009 and February 2010. 

After 2020, the trends rise to between 6.7\% and 7.5\%, a level last reached in 1981. The forward-looking trends rise first and peak in March 2021, at 7.5\% in the attention architecture and 7.2\% in the encoder. The current trends peak at 7.1\% in November 2021 for the attention architecture and at 6.7\% in March 2022 for the encoder. Since their peaks, the trends have fallen back. In December 2025, the two current trends are at 2.7\%, and the forward-looking trends at 2.4\% (encoder) and 2.2\% (attention).

The forward-looking trend therefore turns eight to twelve months before the current trend, and more than a year before inflation itself. When it peaked, in March 2021, 12-month CPI inflation was 2.6\% and still rising, and inflation peaked only in June 2022, at about 9\%; by the time the current trends peaked, the forward-looking trends were already falling. This lead comes from forward-looking nature of the Central Banker's loss, which fits the trend of each month to the inflation of the following twelve months. The lead is therefore a property of the full-sample estimates --- we will evaluate the real-time performance later (Section~\ref{subsec:oos_realtime}).

The two architectures give similar trends, with a mean absolute difference between them of 0.17 percentage points for the current trend and 0.18 for the forward-looking trend. The attention architecture gives the smoother trend: the standard deviation of its monthly changes is 0.23 percentage points, against 0.28 for the encoder, under the Econometrician's loss, and 0.14 against 0.18 under the Central Banker's loss, and its 80\% bands are also narrower. The reason is that the attention layer averages the indicators over past months with similar conditions, which smooths the inputs of the encoder.

How far back the attention mechanism looks depends on its estimated parameters (Section~\ref{subsec:attention}). The decay is between 0.68 and 0.71 under the Econometrician's loss and between 0.77 and 0.80 under the Central Banker's loss, with little variation across the five networks of each estimate. For a given similarity, the weight of a past month is therefore halved after about two months for the current trend and about three months for the forward-looking trend, and a month one year back receives between 1 and 7\% of the weight of the current month, so the attention averages mainly the indicators of the past few months. The sharpness is between 2.6 and 3.0 for the current trend and between 0.48 and 0.65 for the forward-looking trend, so the similarity of the indicators shapes the weights much more for the current trend. 

\subsection{The Volatility of Inflation Surprises}\label{subsec:volatility_estimates}

\begin{figure}[htbp]
    \centering
    \includegraphics[width=\textwidth]{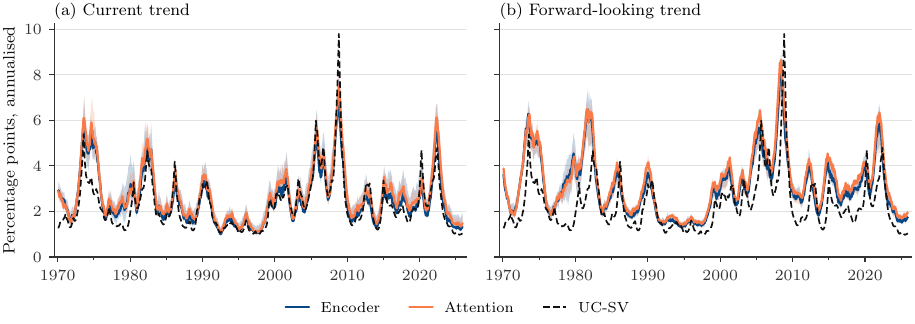}
    \caption{\textbf{The volatility of inflation surprises.} Standard deviation of inflation surprises, $\sigma_t\sqrt{\nu/(\nu-2)}$, in the architectures with time-varying volatility (percentage points, annualised): (a) under the Econometrician's loss, of the error of month $t$; (b) under the Central Banker's loss, of the one-month-ahead forecast error. Encoder in blue and attention architecture in orange, with their 80\% MC-dropout bands. The dashed line is the posterior mean of the standard deviation of the transitory component of the UC-SV model of \citet{StockWatson2007}, estimated on monthly CPI inflation from February 1969 to December 2025.}
    \label{fig:volatility_estimates}
\end{figure}

Figure~\ref{fig:volatility_estimates} shows the volatility of inflation surprises in the two architectures with time-varying volatility. It is the standard deviation of the errors, $\sigma_t\sqrt{\nu/(\nu-2)}$: under the Econometrician's loss, that of the error of month $t$, and under the Central Banker's loss, that of the one-month-ahead forecast error. As a reference, the figure shows the standard deviation of the transitory component of the UC-SV model of \citet{StockWatson2007}, estimated on the same inflation data.

The volatility is high in the mid-1970s, in the early 1980s, in 2008 and in 2021--22, and it is lowest in the 1990s. It follows the volatility of the UC-SV model closely, more so under the Econometrician's loss than under the Central Banker's loss, whose volatility is that of the one-month-ahead forecast errors (panel b).

The volatility state also moves more smoothly than the volatility of the UC-SV model. Its speed limit, $\bar{\xi}$ a month, rules out the one-month spikes of the UC-SV estimate, such as those of 1986 and of April 2020, and it keeps the peak of 2008 at 6.6 to 7.5 percentage points under the Econometrician's loss, against almost 10 for the UC-SV model. Under the Central Banker's loss, the volatility is higher than under the Econometrician's loss in the 2000s and 2010s, and it rises earlier in 2008. In the periods of high volatility, the months carry less weight in the estimation of the trend, and the predictive densities of inflation widen.
 \section{The Drivers of the Trend}
\label{sec:drivers}

As discussed above, in Deep-$\pi^*$, the innovations to the trend are a function of observed indicators, so each movement of the trend can be traced back to them. This section measures how much each group of indicators contributes to the trend in every month. We use these contributions to describe what drove the trend after the pandemic (Section~\ref{subsec:drivers_2019}), what drove the trend in seven episodes of US inflation history (Section~\ref{subsec:episodes}), and how far the contributions depart from those of a linear model (Section~\ref{subsec:nonlinearity}).

\paragraph{Shapley values.} The innovation of a month is a nonlinear function of 122 indicators, so its movements cannot be split among the indicators by reading off coefficients, as in a linear model. We split them instead with Shapley values \citep{Shapley1953}, the standard way of sharing a joint outcome among the players of a cooperative game. Here the players are nine groups of indicators, and the outcome is the innovation of month $t$. For any set $S$ of groups, let $u_t(S)$ be the innovation that the network computes when the indicators of the groups in $S$ take their observed values and those of the other groups are set to their sample medians, which are zero in the standardised data and serve as the reference values. The Shapley value of group $j$ is the change in the innovation when $j$ is added to a set of groups, averaged over all the orders in which the groups can be added:
\begin{equation}\label{eq:shapley}
    \psi_{j,t} = \sum_{S \subseteq \mathcal{G} \setminus \{j\}} \frac{|S|!\,(G - |S| - 1)!}{G!} \Big[ u_t\big(S \cup \{j\}\big) - u_t(S) \Big],
\end{equation}
where $\mathcal{G}$ is the set of the $G = 9$ groups, and the weight of each set $S$ is the probability that exactly the groups in $S$ come before $j$ in a random order. Two properties make Shapley values suited to our purpose. First, they add up: in every month, the values of the nine groups sum to the innovation minus its value with all indicators at their medians. Second, a group that does not change the innovation receives a value of zero, and two groups that play the same role receive the same value. With nine groups, the network is evaluated on all $2^9 = 512$ sets of groups, with dropout switched off. The Shapley values are therefore exact, unlike the SHAP estimators of \citet{Lundberg2017}, such as Kernel SHAP, which approximate them when the players are many individual indicators. In the attention architecture, a group outside $S$ is set to its median in the current month and in all past months, so it also drops out of the similarity that sets the attention weights.

\paragraph{Contributions to the trend.} The trend is an exponentially weighted moving average of the innovations, $\pi_t^* = \theta^t \pi_0^* + (1-\theta) \sum_{s=1}^{t} \theta^{t-s} \tau_s$ (Section~\ref{subsec:model_spec}). Because this average is linear in the innovations, the Shapley values carry over to the trend: the contribution of group $j$ to the trend in month $t$ is the same weighted average of the group's Shapley values,
\begin{equation}\label{eq:filtered_attribution}
    \Psi_{j,t} = (1 - \theta) \sum_{s=1}^{t} \theta^{t-s} \, \psi_{j,s},
\end{equation}
and in every month the trend equals the sum of the contributions of the nine groups plus the path that the trend would follow with all indicators at their medians. Similar to historical shock decompositions in SVAR models, a contribution therefore measures how much higher the trend is in month $t$ because of the values that the indicators of a group took up to that month. A group that pushes the innovation in the same direction for many months builds up a large contribution, whereas a push that lasts a single month adds little and fades slowly, at the rate $\theta$. The contributions of an estimate are the averages of those of its five networks, and they add up to its trend.

The nine groups are based on the classification of the FRED-MD database \citep{McCrackenNg2016}: real activity, employment, housing, orders and inventories, money and credit, stock market, interest and exchange rates, prices, and other indicators. Table~\ref{tab:variable_groups} in the Appendix lists them, with the two differences from that classification. In the figures and tables that follow, orders and inventories, the stock market and the other indicators are shown together as Other.

\subsection{The Drivers of the Post-Pandemic Period Trend}\label{subsec:drivers_2019}

\begin{figure}[htbp]
    \centering
    \includegraphics[width=\textwidth]{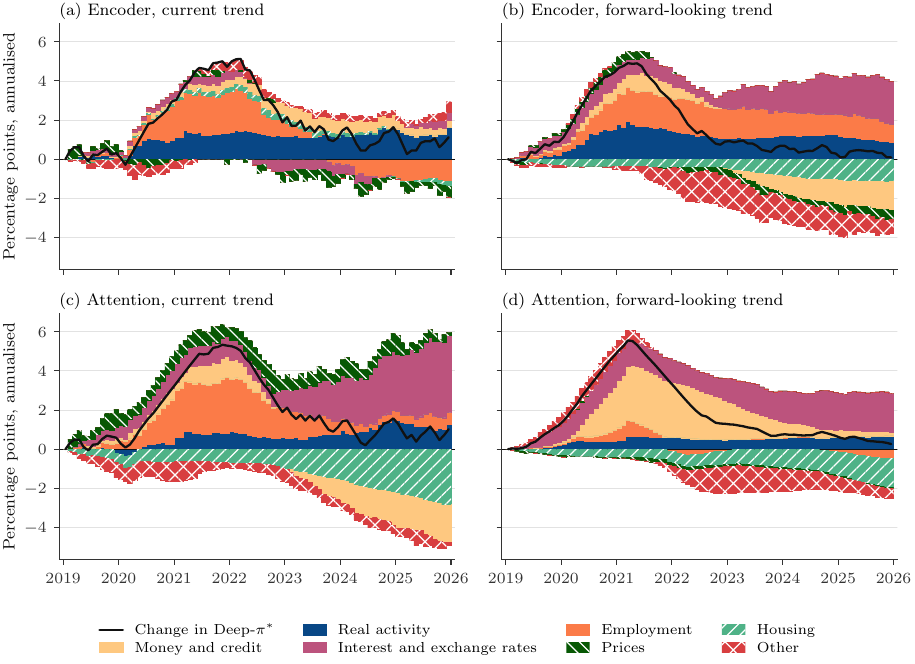}
    \caption{\textbf{The drivers of the trends since January 2019.} Contributions of the groups of indicators to the change in Deep-$\pi^*$ since January 2019 (exact group Shapley values; percentage points, annualised), with time-varying volatility: the encoder (top) and the attention architecture (bottom), under the Econometrician's loss (left) and the Central Banker's loss (right). Orders and inventories, the stock market and the other indicators are shown together as Other. The black line is the change in Deep-$\pi^*$. The path with all indicators at their medians changes by at most 0.13 percentage points after January 2019 and is not drawn.}
    \label{fig:drivers_2019}
\end{figure}

We start by looking at the drivers of the estimated trends in the post-pandemic period. Figure~\ref{fig:drivers_2019} decomposes the change in each trend since January 2019 into the contributions of the groups of indicators, for the encoder and the attention architecture. Table~\ref{tab:drivers_2019_2022} reports, for all eight estimates, the contributions to the change from January 2019 to two dates: the peak of each trend, and June 2022, when 12-month CPI inflation peaked. We report both because the forward-looking trends peaked in March 2021, eight to twelve months before the current trends, so part of their rise had already reversed by June 2022. Three results stand out. The labour market drove the rise of the current trend in both architectures. From the start of the pandemic, real activity, employment, and money and credit drove the rise of the forward-looking trend, with weights that differ between the architectures. Finally, the cooling of the labour market and the contraction of money and credit drove the fall of the trends after their peaks.

\begin{table}[htbp]
    \centering
    \small
    \caption{Contributions to the rise of the trends since January 2019}
    \label{tab:drivers_2019_2022}
    \begin{tabular}{lrrrr}
\toprule
 & \multicolumn{2}{c}{Encoder} & \multicolumn{2}{c}{Attention} \\
\cmidrule(lr){2-3} \cmidrule(lr){4-5}
 & Constant & Time-varying & Constant & Time-varying \\
\midrule
\multicolumn{5}{l}{\textit{Current trend: January 2019 to June 2022}} \\
Real activity & $1.77$ & $1.38$ & $0.69$ & $0.64$ \\
Employment & $1.99$ & $1.62$ & $2.76$ & $2.28$ \\
Housing & $0.53$ & $0.26$ & $-1.39$ & $-0.73$ \\
Money and credit & $0.08$ & $0.62$ & $0.63$ & $0.58$ \\
Interest and exchange rates & $-0.09$ & $-0.29$ & $1.00$ & $0.67$ \\
Prices & $-0.23$ & $0.01$ & $1.64$ & $1.01$ \\
Other & $0.14$ & $0.38$ & $-0.94$ & $-0.38$ \\
\addlinespace[3pt]
Total & $4.37$ & $4.06$ & $4.39$ & $4.04$ \\
\midrule
\multicolumn{5}{l}{\textit{Current trend: January 2019 to its peak}} \\
Month of the peak & Mar 2022 & Mar 2022 & Jan 2022 & Nov 2021 \\
Real activity & $1.84$ & $1.45$ & $1.00$ & $0.83$ \\
Employment & $2.55$ & $2.08$ & $3.16$ & $2.76$ \\
Housing & $0.48$ & $0.23$ & $-1.26$ & $-0.69$ \\
Money and credit & $-0.11$ & $0.49$ & $1.27$ & $1.03$ \\
Interest and exchange rates & $0.46$ & $0.24$ & $1.17$ & $1.02$ \\
Prices & $0.04$ & $0.08$ & $1.25$ & $0.70$ \\
Other & $0.34$ & $0.49$ & $-0.47$ & $-0.30$ \\
\addlinespace[3pt]
Total & $5.77$ & $5.12$ & $6.12$ & $5.34$ \\
\midrule
\multicolumn{5}{l}{\textit{Forward-looking trend: January 2019 to June 2022}} \\
Real activity & $0.75$ & $1.18$ & $1.10$ & $0.52$ \\
Employment & $1.19$ & $1.66$ & $-1.19$ & $-0.27$ \\
Housing & $-0.51$ & $-0.42$ & $-0.82$ & $-0.63$ \\
Money and credit & $0.84$ & $0.32$ & $2.46$ & $2.54$ \\
Interest and exchange rates & $0.68$ & $0.40$ & $1.44$ & $0.96$ \\
Prices & $-0.38$ & $-0.24$ & $-0.18$ & $-0.16$ \\
Other & $-1.61$ & $-1.52$ & $-1.50$ & $-0.98$ \\
\addlinespace[3pt]
Total & $0.89$ & $1.32$ & $1.23$ & $1.91$ \\
\midrule
\multicolumn{5}{l}{\textit{Forward-looking trend: January 2019 to its peak}} \\
Month of the peak & Jun 2021 & Mar 2021 & Apr 2021 & Mar 2021 \\
Real activity & $1.40$ & $1.95$ & $1.40$ & $0.66$ \\
Employment & $1.84$ & $1.73$ & $0.90$ & $0.81$ \\
Housing & $-0.41$ & $-0.37$ & $-0.44$ & $-0.41$ \\
Money and credit & $1.76$ & $0.66$ & $2.65$ & $2.78$ \\
Interest and exchange rates & $0.98$ & $0.78$ & $1.44$ & $1.30$ \\
Prices & $0.31$ & $0.38$ & $-0.10$ & $-0.07$ \\
Other & $-0.28$ & $-0.17$ & $0.33$ & $0.53$ \\
\addlinespace[3pt]
Total & $5.54$ & $4.91$ & $6.13$ & $5.56$ \\
\bottomrule
\end{tabular}
 
    \medskip
    \begin{minipage}{0.95\textwidth}
        \footnotesize \textit{Note:} Contributions of the groups of indicators to the change in Deep-$\pi^*$ from January 2019 to June 2022, and from January 2019 to the peak of each trend (exact group Shapley values; percentage points, annualised), in the eight estimates. The peak is the highest value of the trend between January 2020 and December 2023. Constant and Time-varying refer to the volatility of the inflation equation. Total is the change in Deep-$\pi^*$, which also includes the small change of the path with all indicators at their medians.
    \end{minipage}
\end{table}

\paragraph{Decomposition of the current trend.} The labour market drove the rise of the current trend in both architectures. With time-varying volatility, employment contributes 2.08 of the 5.12 percentage points of the encoder's rise from January 2019 to its peak, and 2.76 of the 5.34 points of the attention architecture's. Within the labour market, payroll employment and hours and the help-wanted indices, which measure vacancies, add to the rise in both architectures (Table~\ref{tab:robust_finer} in the Appendix): 1.31 and 1.01 points in the encoder, and 1.09 and 0.66 points in the attention architecture, in which unemployment and claims add a further 1.39 points. These indicators describe the tight labour market of 2021 and 2022, when vacancies were very high relative to the number of unemployed \citep{DomashSummers2022, BallLeighMishra2022}. The other groups add less. Real activity contributes 1.45 points in the encoder and 0.83 points in the attention architecture, in which money and credit and interest and exchange rates add about 1 point each.

\paragraph{Decomposition of the forward-looking trend.} The two architectures attribute the rise of the forward-looking trend to the same groups, but they weigh them differently. In the encoder, real activity and employment contribute most, 1.95 and 1.73 of the 4.91 points up to its peak, followed by interest and exchange rates with 0.78 points and money and credit with 0.66 points. In the attention architecture, money and credit contributes 2.78 of the 5.56 points, interest and exchange rates 1.30 and employment 0.81. Part of this rise came before the pandemic: the two forward-looking trends had already risen during 2019, mostly with interest and exchange rates, which by February 2020 had made most of their contribution up to the peaks. From the first months of the pandemic, real activity, employment, and money and credit push the trends up (Figure~\ref{fig:drivers_2019}, panels b and d), and the contribution of money and credit to the attention architecture's trend keeps growing until the middle of 2021.

\paragraph{Why the two architectures differ.} The difference comes from how the two architectures use the past lags of the indicators. In the spring of 2020, money and credit surged within a few months: the monetary aggregates and bank reserves expanded sharply, and firms drew on their credit lines. However, inflation did not rise concurrently. The encoder reads the indicators of each month on their own, so it registers this surge only in the months in which it happened. The attention architecture averages the indicators over recent months, and under the Central Banker's loss its memory is the longer one (Section~\ref{subsec:trends}), so the surge stays in its inputs for several months, during which it keeps adding to the inflation trend. We can measure how much of the contribution of money and credit comes from these past months. Keeping the estimated attention weights, we recompute the trend with only the current month of the money and credit indicators in the weighted average. Their contribution to the rise of the forward-looking trend then falls from 2.78 to 1.31 percentage points (Table~\ref{tab:robust_memory} in the Appendix), so about half of it comes from the surge of the preceding months, and the remaining 1.31 points are still about twice the 0.66 points of the encoder. Real activity shows the opposite pattern. It collapsed in April 2020 and rebounded in the following months, and in an average over recent months the collapse and the rebound partly offset each other. Its contribution to the attention architecture's forward-looking trend therefore stays at about 0.4 points throughout 2020, while in the encoder it builds up with the recovery, to about 1.5 points in the last months of the year. The two architectures thus give a different timing to the same events: the attention architecture lets the monetary expansion of 2020 act with a lag of several months, in the spirit of the lags of monetary policy of \citet{Friedman1961}, whereas the encoder lets the recovery of activity act on impact. Out of sample, neither architecture forecasts inflation better than the other at the 5\% level in the post-pandemic episode (Table~\ref{tab:oos_encoder_attention} in Appendix~\ref{app:out_of_sample}).

\paragraph{The fall after the peaks.} After their peaks, the trends fall mainly with employment and with money and credit, as the labour market cooled and the money stock contracted in 2022 and 2023. Interest and exchange rates, by contrast, push the trends up, in particular from 2023: from January 2023 to December 2025, their contribution rises in all four estimates, mostly through the spreads of market interest rates over the federal funds rate. These spreads fell as the federal funds rate rose above long-term government bond yields and close to corporate bond yields in 2023 and 2024. 

These results bear on the distinction between demand and supply that matters for monetary policy. The groups that drove the rise of both trends after the pandemic, the labour market, real activity, and money and credit, are those usually associated with aggregate demand and monetary conditions. The prices group, which holds producer, oil and commodity prices, average hourly earnings and the consumer prices of the previous month, contributed little: from January 2019 to the peaks of the trends with time-varying volatility, between $-0.1$ and 0.7 of the 4.9 to 5.6 percentage points of their rise, in a period in which energy and commodity prices surged (Table~\ref{tab:drivers_2019_2022}). The trend therefore looked through most of the movements in prices that are usually attributed to supply shocks, whose effects on inflation tend to be temporary. This reading is consistent with \citet{BernankeBlanchard2025}, who find that the effects of a tight labour market on US inflation are more persistent than those of shocks to product prices. The groups of indicators are not structural shocks, however: employment, for example, reflects labour supply as well as labour demand, and the prices group also holds average hourly earnings, so identifying the demand and supply shocks behind the contributions would require further structure.

Finally, the time-varying volatility changes the attribution little. Across the four pairs of estimates with constant and time-varying volatility in Table~\ref{tab:drivers_2019_2022}, the contribution of any group to the rise of the trends to their peaks changes by at most 0.74 percentage points, with one exception, money and credit in the encoder's forward-looking trend, which contributes 1.76 points with constant volatility and 0.66 with time-varying volatility. The largest contributor to the rise of each trend is the same in three of the four pairs; in the encoder's forward-looking trend, it is employment with constant volatility and real activity with time-varying volatility. The reference values of the Shapley values matter little as well: with the indicators at their 2010--2019 averages instead of their medians, no contribution to the rise of the trends to their peaks changes by more than 0.52 points (Table~\ref{tab:robust_reference}).

\subsection{The Drivers in Seven Episodes}\label{subsec:episodes}

\begin{figure}[htbp]
    \centering
    \includegraphics[width=\textwidth]{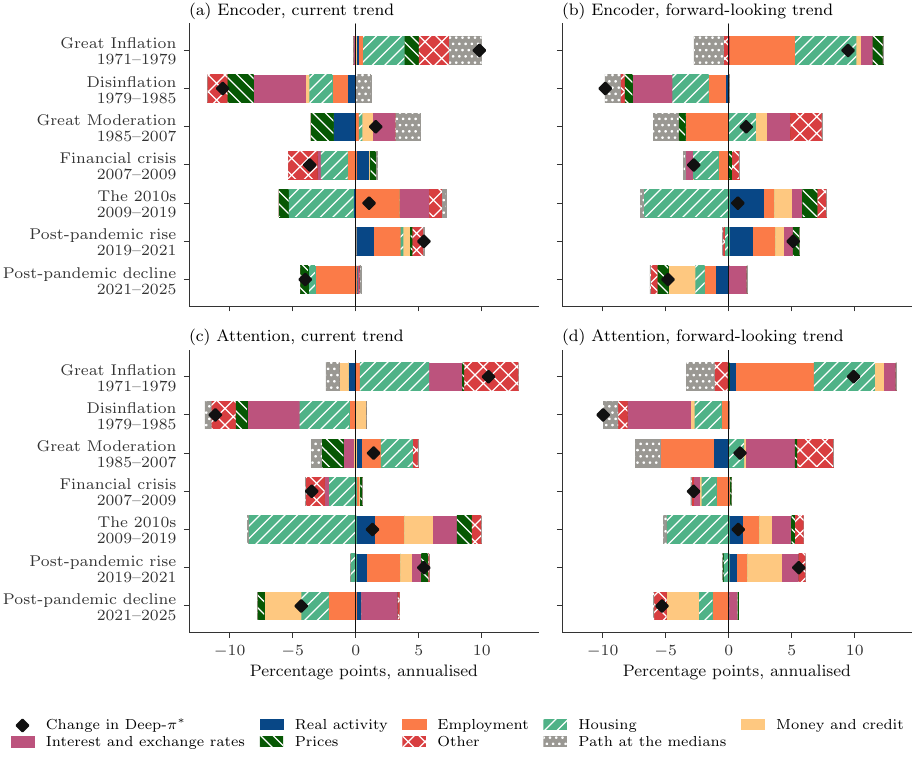}
    \caption{\textbf{The drivers of the trends in seven episodes.} Change in Deep-$\pi^*$ over each episode (diamonds) and the contributions of the groups of indicators to it (exact group Shapley values; percentage points, annualised), with time-varying volatility: the encoder (top) and the attention architecture (bottom), under the Econometrician's loss (left) and the Central Banker's loss (right). The years under each episode are those of the median turning point across the four estimates; Table~\ref{tab:turning_points} gives the turning points of each estimate. Path at the medians: the change of the path with all indicators at their medians. Orders and inventories, the stock market and the other indicators are shown together as Other.}
    \label{fig:episodes}
\end{figure}

We turn our focus now to other inflation periods. Figure~\ref{fig:episodes} decomposes the change in each trend over seven episodes of US inflation history: the Great Inflation, the disinflation of the early 1980s, the Great Moderation, the financial crisis, the 2010s, and the rise and decline of the trend after the pandemic. Each episode runs between two turning points of the trend, which we date as the highest or lowest value of the trend within a window of years (Table~\ref{tab:turning_points} in the Appendix).

In the Great Inflation, the trends rise by 9.5 to 10.5 percentage points. Housing contributes between 3.3 and 5.5 points in all four estimates. Employment contributes 5.2 and 6.2 points to the forward-looking trends and less than half a point to the current trends, which rise instead with the indicators shown as Other (2.4 and 4.3 points), almost all of it from orders and inventories. In the disinflation that follows, the trends fall by 10 to 11 points. Interest and exchange rates are the largest contributor in all four estimates, with between $-3.2$ and $-5.0$ points, and housing contributes between $-1.8$ and $-3.9$. Money and credit contributes less than one point to either episode.

The disinflation also shows how the forward-looking trend reads interest rates. The forward-looking trend peaked in early 1979, a year before inflation and about two years before the federal funds rate, and it was already falling when the Federal Reserve under Paul Volcker tightened monetary policy from October 1979 \citep{GoodfriendKing2005}. From its peak to the end of 1980, interest and exchange rates were the largest contributor to its fall, most of it from the rise of long-term rates, and money and credit came next. In the forward-looking trend, rising long-term rates lower the trend: they did so during the Volcker tightening, and again in 2022. 

Over the Great Moderation and the 2010s, the trends change by less than 2 points, while large contributions of opposite sign offset each other. In the 2010s, for example, housing subtracts 4.9 to 8.5 points, and the other groups together add a similar amount.

In the financial crisis, the trends fall by 2.8 to 3.6 points. The fall comes from several groups: housing (up to $-2.1$ points), employment (up to $-1.0$) and, in the current trends, the indicators shown as Other ($-2.3$ and $-1.5$), mainly the stock market and orders and inventories. The contribution of employment is modest, although the unemployment rate doubled between the end of 2007 and the end of 2009, from 5\% to 10\%. This is in line with the missing disinflation of the Great Recession studied by \citet{CoibionGorodnichenko2015}.

After their post-pandemic peaks, the trends fall by 4.0 to 5.3 points up to December 2025. Employment accounts for most of the fall of the encoder's current trend ($-3.2$ points), and money and credit is the largest contributor to the fall of the forward-looking trends ($-2.2$ and $-2.6$). In the attention architecture's current trend, employment and housing contribute $-2.2$ points each, and money and credit $-3.0$. Interest and exchange rates push the trends up over the same period, by 0.2 to 2.9 points, as discussed in Section~\ref{subsec:drivers_2019}.

However, we should emphasise that the contributions are merely statistical attributions. They show how the network maps the indicators into the trend, and they do not by themselves identify causal effects. They are nonetheless economically coherent across episodes. Strong housing activity in the 1970s and the tight labour market and monetary expansion of 2020 and 2021 raised the trend, and the cooling of the labour market and the contraction of money brought it down. Interest rates show how the two trends use the same indicators. For the forward-looking trend, rising long-term rates signal lower inflation ahead, in line with a tightening of monetary conditions, as in the Volcker years and in 2022. 

\subsection{Are There Nonlinearities?}\label{subsec:nonlinearity}

The use of neural networks also allows us to understand if there are nonlinearities in the drivers of the trends. First, the contribution of a group can be a nonlinear function of its own indicators: a tight labour market, for example, could raise the trend more than proportionally. Second, the contribution of a group can depend on the values of the other groups: the effect of money growth on the trend could depend on the state of the labour market. To quantify the possible nonlinearities, we use two simple measures that come from the sets of groups already evaluated for the Shapley values.

The \textit{nonlinearity share} measures how far the Shapley value of a group is from a linear function of its own indicators. It is one minus the adjusted $R^2$ of a regression of the group's Shapley value, $\psi_{j,t}$, on the group's indicators over the full sample, and it is zero when the contribution of the group is a linear function of them. For the attention architecture, the regressors are the outputs of the attention layer for the group's indicators, which are the inputs of its encoder. Because the Shapley value also contains the group's interactions with the other groups, the nonlinearity share picks up both departures from a linear model.

The \textit{interaction share} isolates the second departure. It splits the Shapley value into two parts: the main effect of the group, $m_{j,t} = u_t(\{j\}) - u_t(\emptyset)$, which is the change in the innovation when only that group takes its observed values, and the rest, $\psi_{j,t} - m_{j,t}$, which comes from the group's interactions with the other groups. The share compares the size of the two parts,
\begin{equation}\label{eq:interaction_share}
    \frac{\sum_t \big(\psi_{j,t} - m_{j,t}\big)^2}{\sum_t m_{j,t}^2 + \sum_t \big(\psi_{j,t} - m_{j,t}\big)^2},
\end{equation}
and it lies between zero and one, with zero when the network is additive in the groups. It is close in spirit to the statistic of \citet{FriedmanPopescu2008} for the interactions of one variable with all the others. Finally, the bound on the innovations in equation~\eqref{eq:innovation} is itself a mild nonlinearity, which matters only in months with large innovations. Table~\ref{tab:nonlinearity} reports the two shares for each group, in the four estimates with time-varying volatility.

\begin{table}[htbp]
    \centering
    \small
    \caption{Nonlinearity and interactions of the groups of indicators}
    \label{tab:nonlinearity}
    \begin{tabular}{lrrrr}
\toprule
 & \multicolumn{2}{c}{Current trend} & \multicolumn{2}{c}{Forward-looking trend} \\
\cmidrule(lr){2-3} \cmidrule(lr){4-5}
 & Encoder & Attention & Encoder & Attention \\
\midrule
\multicolumn{5}{l}{\textit{Nonlinearity share}} \\
Real activity & $0.31$ & $0.32$ & $0.39$ & $0.53$ \\
Employment & $0.29$ & $0.39$ & $0.37$ & $0.31$ \\
Housing & $0.17$ & $0.21$ & $0.16$ & $0.26$ \\
Orders and inventories & $0.38$ & $0.42$ & $0.25$ & $0.28$ \\
Money and credit & $0.40$ & $0.40$ & $0.30$ & $0.13$ \\
Stock market & $0.28$ & $0.36$ & $0.26$ & $0.43$ \\
Interest and exchange rates & $0.37$ & $0.43$ & $0.24$ & $0.25$ \\
Prices & $0.36$ & $0.49$ & $0.34$ & $0.35$ \\
Others & $0.54$ & $0.60$ & $0.47$ & $0.78$ \\
\midrule
\multicolumn{5}{l}{\textit{Interaction share}} \\
Real activity & $0.20$ & $0.25$ & $0.20$ & $0.34$ \\
Employment & $0.19$ & $0.24$ & $0.19$ & $0.15$ \\
Housing & $0.23$ & $0.24$ & $0.17$ & $0.18$ \\
Orders and inventories & $0.24$ & $0.25$ & $0.17$ & $0.15$ \\
Money and credit & $0.26$ & $0.26$ & $0.17$ & $0.15$ \\
Stock market & $0.28$ & $0.26$ & $0.18$ & $0.36$ \\
Interest and exchange rates & $0.21$ & $0.27$ & $0.15$ & $0.14$ \\
Prices & $0.11$ & $0.19$ & $0.18$ & $0.22$ \\
Others & $0.30$ & $0.30$ & $0.24$ & $0.36$ \\
\bottomrule
\end{tabular}
 
    \medskip
    \begin{minipage}{0.95\textwidth}
        \footnotesize \textit{Note:} Nonlinearity share (one minus the adjusted $R^2$ of a linear regression of the group's Shapley value on its own inputs) and interaction share, equation~\eqref{eq:interaction_share}, of each group of indicators, averaged over the five networks of each estimate with time-varying volatility.
    \end{minipage}
\end{table}

The departures from a linear model are far from negligible. In the encoder, a linear function of a group's own indicators cannot reproduce between 16 and 40\% of the variation of its Shapley value for eight of the nine groups, and the interactions with the other groups account for between 11 and 28\% of its size. Among the groups with large contributions, real activity, employment, money and credit, interest and exchange rates, and prices all have nonlinearity shares between 0.24 and 0.40. The other indicators have the highest nonlinearity shares in the encoder (0.54 and 0.47), but they contribute little to the trend, and their share varies widely across the five networks of each estimate.

In the attention architecture the shares are of a similar size, between 0.13 and 0.53 for nonlinearity and between 0.14 and 0.36 for interactions for eight of the nine groups. Under the Econometrician's loss, its nonlinearity shares are at least as high as those of the encoder for every group, and under the Central Banker's loss the highest values are those of real activity, 0.53 for nonlinearity and 0.34 for interactions, and the interaction share of the stock market, 0.36. For real activity in that estimate, a linear function of its own inputs therefore misses more than half of the variation of its contribution. These shares come on top of the nonlinearity of the attention layer itself, whose output, which the encoder of this architecture reads, is already a nonlinear function of the indicators through the attention weights.

These nonlinearities are particularly visible in the post-pandemic period, when many indicators moved far outside their usual range, and two of them are worth highlighting. The first is the interaction between employment and money and credit. In 2020 and 2021, both groups pointed in the same direction, towards an overheating economy, but the network does not add up their two signals in full, so their joint push on the trend is smaller than the sum of their separate pushes. Measured with the Shapley interaction index of \citet{GrabischRoubens1999}, which compares the effect of adding the two groups together with the sum of their separate effects, the interaction reduces the rise of the trends to their peaks in all four estimates, and it partly offsets their fall afterwards. The second is the response of the trend to the labour market itself, which was larger than a linear model would imply. In three of the four estimates, the nonlinear part of the main effect of employment, the part of $m_{j,t}$ that a linear regression on the group's own indicators does not capture, adds to the rise of the trends, while it is slightly negative in the encoder's current trend. In other words, the network responded more than proportionally to the large movements of the labour market in 2020 and 2021, in the spirit of the nonlinear Phillips curves of \citet{BenignoEggertsson2023} and \citet{BuckmannPotjagailoSchnattinger2025}, in which a tight labour market raises inflation more than proportionally.

The nonlinearities also matter for the trend as a whole. Estimated with a single linear layer from the indicators to the innovation, Deep-$\pi^*$ gives a trend that follows the same path but is between 2.1 and 2.2 times as volatile from month to month, and its predictive densities are less accurate out of sample (Appendix~\ref{app:linear}).
 \section{Out-of-Sample and Real-Time Evaluation}
\label{sec:out_of_sample}

We now evaluate the performance of the model in out-of-sample forecasting, and we also check how much the trends are revised when new data become available. Estimates of latent variables that look reasonable with hindsight can be revised substantially in real time: \citet{OrphanidesVanNorden2002} show that the revisions of real-time estimates of the US output gap are of the same order of magnitude as the gap itself, and that they come mostly from the difficulty of estimating the trend at the end of the sample. This is a feature of unobserved components models, in which the trend has a stochastic shock of its own: even with the parameters known, later observations revise the estimate of the trend in a month. In Deep-$\pi^*$, by contrast, the trend in a month is a deterministic function of the indicators up to that month, given the parameters, so later data can revise it only when the model is re-estimated and the parameters change. Unless the parameters change, new data only gives an updated estimate of the trend when for the periods in which there is new data---the past trend remains thus unrevised. In this section we estimate the model as a user would have done during the post-pandemic surge of inflation, re-estimating it every month with the data available at the time, and we ask how much the trend was revised, what the model attributed the trend to at the time, and how accurately the model forecast inflation compared with standard benchmarks and with the Survey of Professional Forecasters.

\subsection{Design of the Exercise}\label{subsec:oos_design}

We re-estimate each of the eight trends with the data up to the end of every month from December 2020 to November 2023. All the estimations use the FRED-MD vintage of September 2026, so the exercise leaves out the revisions of the data themselves and is in quasi-real time. 

Throughout the section we distinguish the trend from the forecast of inflation. An estimation with the data up to month $T$ gives the trend in every month up to $T$, and since no indicators of later months are available, its projection $h$ months ahead is $\theta^h \pi^*_T$, which with $\theta = 0.995$ moves only slowly towards the median of inflation in the estimation sample. The forecast of inflation combines this projection with the autoregressive dynamics of the inflation gap, as in Section~\ref{subsec:loss}, so that in the first months it also reflects the deviation of recent inflation from the trend. Figures~\ref{fig:oos_spaghetti} and~\ref{fig:oos_drivers} show the trend, its projection and its drivers, while the tables and Figures~\ref{fig:oos_cumulative_monthly} and~\ref{fig:oos_spf} evaluate the forecasts of inflation, which for short we call the forecasts of the current and of the forward-looking trend.

Moreover, the two loss functions that we consider use the end of an estimation sample differently. Under the Econometrician's loss, every month enters with its own inflation, so the current trend in the last month of the sample is fitted to the inflation of that month. Under the Central Banker's loss, by contrast, a month is evaluated on the inflation of the twelve months that follow it, which at the end of the sample are only partly observed. In an estimation with the data up to December 2020, for instance, the trend of June 2020 enters the loss with its forecasts for July to December 2020, and that of November 2020 with its forecast for December 2020 only (equation~\eqref{eq:loss_forecast}). The forward-looking trend of the last months of a sample therefore rests on the part of the following year that has already been observed, and it is pinned down fully only once that year is complete.\footnote{If the last months of a sample entered the loss only once their twelve outcomes were observed, the network would learn from the indicators of a month a year later. In the post-pandemic episode, the forward-looking trend would then have risen later in 2021 and stayed high for longer in 2022 in real time.}

We evaluate the forecasts of inflation on two objects, both made with the data up to the month of the forecast: the forecast of inflation in the following month, and the average of the forecasts over the next twelve months, whose errors are those of average inflation over the next year. The predictive densities are the Student-$t$ densities of the inflation equation, mixed over 200 MC-dropout passes. With time-varying volatility, the scale of the forecast $h$ months ahead is $\sigma_{t,h}$ of equation~\eqref{eq:scale_h} under the Central Banker's loss, and $\bar{\sigma}\exp\big(\bar{v}\tanh(\theta^h\lambda_t/\bar{v})\big)$, without horizon factors, under the Econometrician's loss; with constant volatility, it is $\sigma$ at every horizon. We report the root mean squared error (RMSE) of the point forecasts, the log score of the densities, and the continuous ranked probability score (CRPS), which measures the distance between the predictive distribution and the outcome in percentage points. For the twelve-month object, the log score and the CRPS are the averages over the twelve monthly forecasts. We test for equal accuracy with the test of \citet{DieboldMariano1995}, applied to the log scores as in \citet{AmisanoGiacomini2007}, to the CRPS and to the squared errors, with standard errors that allow for the serial correlation of the forecast errors \citep{NeweyWest1987}, which is strong for the twelve-month averages of consecutive months.

We compare Deep-$\pi^*$ with three statistical benchmarks, each re-estimated every month: an autoregression with twelve lags and Student-$t$ errors, AR(12); the UC-SV model of \citet{StockWatson2007}, estimated by Gibbs sampling with the precision-based sampler of \citet{ChanJeliazkov2009},\footnote{We use the Python code of Joshua Chan, available at \url{https://github.com/joshuaccchan/bayesian-macroeconometrics}.} whose forecast at every horizon is its trend; and the random walk of \citet{AtkesonOhanian2001}, which forecasts every month of the next year by the average inflation of the past twelve months. We also compare the forecasts over the next twelve months with the median CPI forecast of the Survey of Professional Forecasters \citep[SPF;][]{Croushore1993}.

\subsection{The Trend in Real Time}\label{subsec:oos_realtime}

\begin{figure}[htbp]
    \centering
    \includegraphics[width=\textwidth]{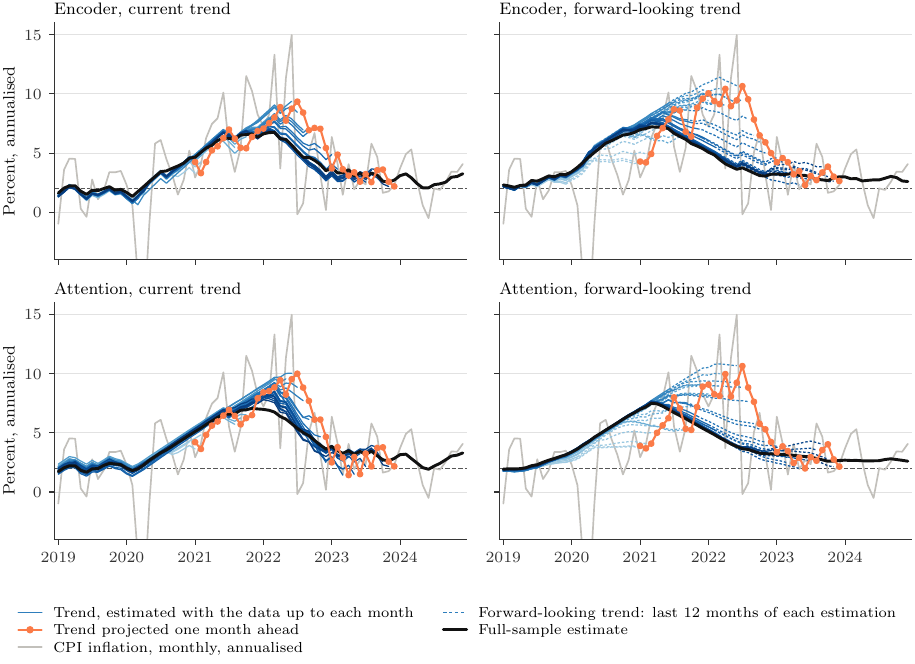}
    \caption{\textbf{The trend re-estimated every month, December 2020 to November 2023.} Deep-$\pi^*$ with time-varying volatility (percent, annualised): the encoder (top) and the attention architecture (bottom), under the Econometrician's loss (left) and the Central Banker's loss (right). Each thin blue line is the trend estimated with the data up to one month, from light (December 2020) to dark (November 2023). For the forward-looking trend, the dashed part of each line covers the last twelve months of its sample, which enter the estimation with the horizons observed by the end of the sample. The orange dots are the trend projected one month ahead from each estimation, plotted in the month after its last data. The black line is the full-sample estimate (five networks), and the grey line is monthly CPI inflation, annualised. The figure shows the trend only; the forecasts of inflation are evaluated in Table~\ref{tab:oos_monthly}.}
    \label{fig:oos_spaghetti}
\end{figure}

Figure~\ref{fig:oos_spaghetti} shows the trend estimated in quasi-real time, from 2019 onwards. Each thin blue line is the trend from one estimation, with the data up to a given month, and the black line is the full-sample estimate. The distance between the two therefore shows how much the trend was revised as the network was re-estimated with later data. The orange dot after each blue line is the trend projected one month ahead.

For the forward-looking trend, the last twelve months of each blue line are dashed, for the reason given in Section~\ref{subsec:oos_design}: the trend of a month is fitted to inflation over the following twelve months, which at the end of the sample have not all been observed yet. In the estimation with the data up to December 2020, for instance, the trend of June 2019 is fitted to inflation from July 2019 to June 2020, all of which was known, so this part of the line is solid. The trend of June 2020, by contrast, is fitted only to inflation from July to December 2020, so this part of the line is dashed: it is still partly a forecast, and later estimations revise it as the rest of the year is observed and new data are available.

The current trend is revised about as much as the UC-SV trend. In real time it follows the full-sample estimate closely during 2021, rises above it from the end of the year and peaks in June 2022, the month in which 12-month CPI inflation peaked, at 9.4\% in the encoder and 10.0\% in the attention architecture, against 6.7\% and 7.1\% in the full-sample estimates (Table~\ref{tab:oos_peaks} in Appendix~\ref{app:out_of_sample}). By the end of 2023 the two are close again, and over the thirty-six months the real-time trend differs from the full-sample estimate by 1.0 and 1.2 percentage points on average, against 1.2 points for the UC-SV trend, which in real time peaks at 10.7\% in March 2022, against 7.3\% in its full-sample estimate.

The revisions of the forward-looking trend are concentrated in the last months of each sample. In the months whose following year had already been observed, the solid parts of the lines, the trend estimated in real time stays close to the full-sample estimate and changes little from one estimation to the next. In the dashed parts, the forward-looking trend is in part a forecast of inflation over a year that is still to come, and it is revised as that year unfolds. In the last month of each sample, the forward-looking trend therefore differs from the full-sample estimate by 2.3 points in the encoder and 2.1 points in the attention architecture on average (Table~\ref{tab:oos_peaks}), and this difference shrinks once the following year has been observed. Section~\ref{subsec:oos_drivers} shows that the large downward revisions of 2022 came mainly from the contribution of the labour market. Since the real-time forward-looking trend is a forecast, the natural way to judge it is by its accuracy, which Section~\ref{subsec:oos_accuracy} evaluates.

In real time, the forward-looking trend pointed in the right direction throughout the episode. In December 2020 it was already at 4.3\% in the encoder and 3.9\% in the attention architecture, when 12-month CPI inflation was 1.3\%, and it stayed above 12-month inflation in every month of the rise, up to a peak of 10.7\% in June 2022 in both architectures, the month in which 12-month CPI inflation peaked. From August 2022, while 12-month inflation was still close to 8\%, the forward-looking trend fell below it, to 4.2\% and 3.4\% in December 2022, as the lower monthly inflation of the second half of 2022 entered the estimation, and it ended at 2.6\% and 2.1\% in November 2023.

\subsection{The Drivers in Real Time}\label{subsec:oos_drivers}

\begin{figure}[htbp]
    \centering
    \includegraphics[width=\textwidth]{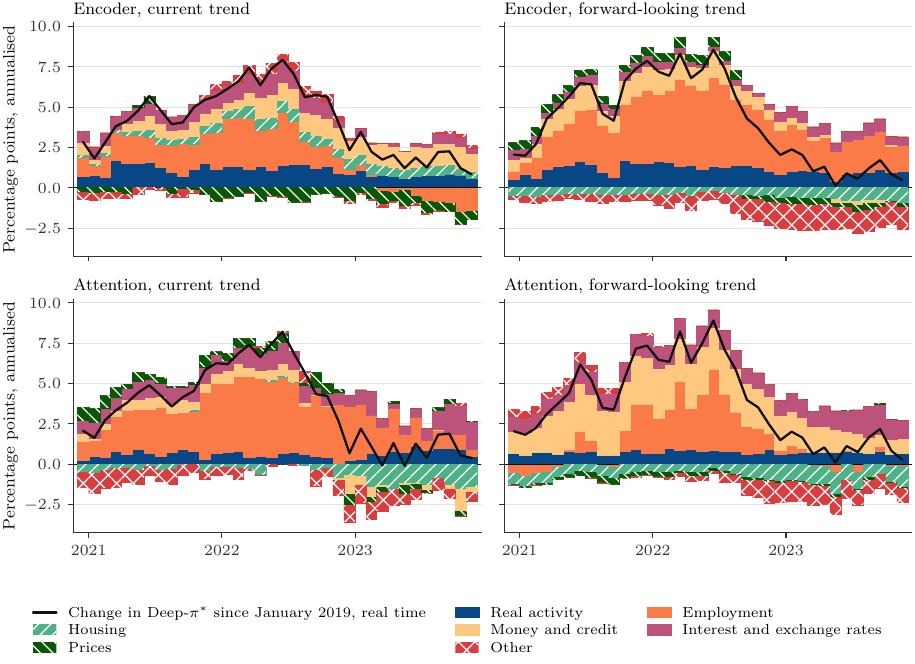}
    \caption{\textbf{The drivers of the trends in real time, December 2020 to November 2023.} Contributions of the groups of indicators to the change in Deep-$\pi^*$ since January 2019 (exact group Shapley values; percentage points, annualised), with time-varying volatility: the encoder (top) and the attention architecture (bottom), under the Econometrician's loss (left) and the Central Banker's loss (right). Each bar is one month, from the network estimated with the data up to that month. The black line is the change in the trend in real time. The path with all indicators at their medians changes by at most 0.22 percentage points in these runs and is not drawn.}
    \label{fig:oos_drivers}
\end{figure}

Because the decompositions of Section~\ref{sec:drivers} can be computed for every estimation, we can also see what the model attributed the trend to at the time. Figure~\ref{fig:oos_drivers} shows, for each month from December 2020 to November 2023, the contributions of the groups of indicators to the change in the trend since January 2019, from the network estimated with the data up to that month.

In real time, the rise of the forward-looking trend came mainly from the labour market and from money and credit. In the encoder, employment contributed 2.1 percentage points by March 2021 and 4.5 points by December 2021, while real activity and money and credit contributed about 1.5 points each. In the attention architecture, money and credit pushed the trend up from the end of 2020, by 2.3 points in March 2021 and by about 2.9 points through the rest of the year, and employment followed, with 3.0 points in December 2021. At the peak, in June 2022, employment contributed 5.6 points in the encoder and 5.0 points in the attention architecture (Table~\ref{tab:oos_turn}). The current trend also rose mainly with employment in real time, which contributed 3.2 and 4.7 points to it in June 2022.

\begin{table}[htbp]
    \centering
    \small
    \caption{The turn of the forward-looking trend in real time}
    \label{tab:oos_turn}
    \resizebox{\textwidth}{!}{\begin{tabular}{lrrrrrr}
\toprule
 & \multicolumn{3}{c}{Encoder} & \multicolumn{3}{c}{Attention} \\
\cmidrule(lr){2-4} \cmidrule(lr){5-7}
 & Peak (Jun 2022) & June 2023 & Change & Peak (Jun 2022) & June 2023 & Change \\
\midrule
Real activity & $1.33$ & $0.82$ & $-0.50$ & $0.86$ & $0.82$ & $-0.05$ \\
Employment & $5.55$ & $2.06$ & $-3.50$ & $5.03$ & $0.05$ & $-4.98$ \\
Housing & $-0.24$ & $-0.75$ & $-0.52$ & $-0.29$ & $-1.04$ & $-0.75$ \\
Money and credit & $1.63$ & $-0.24$ & $-1.87$ & $2.63$ & $1.17$ & $-1.47$ \\
Interest and exchange rates & $0.26$ & $0.64$ & $0.39$ & $1.05$ & $1.32$ & $0.27$ \\
Prices & $0.56$ & $-0.25$ & $-0.81$ & $-0.14$ & $0.00$ & $0.14$ \\
Other & $-0.45$ & $-1.29$ & $-0.84$ & $-0.15$ & $-1.08$ & $-0.94$ \\
Total & $8.56$ & $0.89$ & $-7.67$ & $8.91$ & $1.13$ & $-7.78$ \\
\bottomrule
\end{tabular}
 }

    \medskip
    \begin{minipage}{0.95\textwidth}
        \footnotesize \textit{Note:} Contributions of the groups of indicators to the change in the forward-looking trend since January 2019 (exact group Shapley values; percentage points, annualised), with time-varying volatility, from the networks estimated with the data up to the month of the real-time peak of the trend and up to June 2023, and the change between the two. Total: the change in the trend since January 2019, which also includes the small change of the path with all indicators at their medians. Other: orders and inventories, the stock market and the other indicators.
    \end{minipage}
\end{table}

The same groups account for most of the turn of the trend. From the estimation of June 2022 to that of June 2023, the rise of the forward-looking trend since January 2019 fell from 8.6 to 0.9 percentage points in the encoder and from 8.9 to 1.1 points in the attention architecture. Of this fall of 7.7 and 7.8 points, employment accounts for 3.5 and 5.0 points, money and credit for 1.9 and 1.5 points, and the indicators shown as Other for 0.8 and 0.9 points. Housing accounts for a further 0.5 and 0.7 points, and in the encoder prices and real activity for 0.8 and 0.5 points. Interest and exchange rates, by contrast, added 0.4 and 0.3 points to the trend over the same period.

The smaller contribution of employment also explains the revisions of the forward-looking trend in 2022 (Section~\ref{subsec:oos_realtime}). Estimated with the data up to mid-2022, the network had learned from 2021 that a tightening labour market was followed by rising inflation, and it read the very tight labour market of mid-2022 as a sign that inflation would stay high. Inflation in the following months turned out lower, although the labour market remained tight, so the following estimations gave less weight to employment, which lowered both the trend in real time and the trend of the months before. The episode thus taught the network that the link between the labour market and future inflation was weaker than in 2021, for reasons that the model does not separate, such as the easing of supply bottlenecks and the tightening of monetary policy.

\subsection{Forecast Accuracy}\label{subsec:oos_accuracy}

We now turn to the forecasts of inflation. Table~\ref{tab:oos_monthly} compares them with those of the benchmarks. One month ahead, both trends forecast inflation more accurately than UC-SV and the random walk, and about as accurately as the autoregression. Their log scores exceed those of UC-SV by 0.13 to 0.19 and those of the random walk by 0.24 to 0.30, and their CRPS is 11 to 13\% below that of UC-SV, differences that are all significant at the 10\% level and most of them at the 5\% level. Against the autoregression, the log scores are higher by 0.09 to 0.15 and the RMSE is up to 4\% lower, but these differences are not significant.

\begin{table}[htbp]
    \centering
    \small
    \caption{Forecast accuracy against the benchmarks, 2021--2023}
    \label{tab:oos_monthly}
    \resizebox{\textwidth}{!}{\begin{tabular}{lrrrrrrrrr}
\toprule
 & \multicolumn{3}{c}{Log-score difference} & \multicolumn{3}{c}{RMSE ratio} & \multicolumn{3}{c}{CRPS ratio} \\
\cmidrule(lr){2-4} \cmidrule(lr){5-7} \cmidrule(lr){8-10}
 & AR(12) & UC-SV & Random walk & AR(12) & UC-SV & Random walk & AR(12) & UC-SV & Random walk \\
\midrule
\multicolumn{10}{l}{\textit{One month ahead}} \\
Current trend, encoder & $0.09$ & $0.13$$^{*}$ & $0.24$$^{**}$ & $0.98$ & $0.93$$^{**}$ & $0.90$ & $0.96$ & $0.89$$^{**}$ & $0.83$$^{**}$ \\
Current trend, attention & $0.11$ & $0.15$$^{**}$ & $0.27$$^{**}$ & $0.96$ & $0.91$$^{**}$ & $0.88$ & $0.93$ & $0.87$$^{***}$ & $0.81$$^{**}$ \\
Forward-looking trend, encoder & $0.15$ & $0.19$$^{**}$ & $0.30$$^{**}$ & $0.98$ & $0.93$ & $0.90$ & $0.93$ & $0.87$$^{**}$ & $0.81$$^{**}$ \\
Forward-looking trend, attention & $0.11$ & $0.15$$^{*}$ & $0.26$$^{**}$ & $1.00$ & $0.95$ & $0.92$ & $0.96$ & $0.89$$^{*}$ & $0.83$$^{*}$ \\
\midrule
\multicolumn{10}{l}{\textit{Next twelve months}} \\
Current trend, encoder & $0.00$ & $0.01$ & $0.15$$^{***}$ & $1.06$ & $0.83$$^{***}$ & $0.75$$^{**}$ & $1.02$ & $0.93$$^{**}$ & $0.85$$^{**}$ \\
Current trend, attention & $-0.03$ & $-0.02$ & $0.11$ & $1.04$ & $0.82$$^{***}$ & $0.74$$^{**}$ & $1.02$ & $0.93$ & $0.85$$^{**}$ \\
Forward-looking trend, encoder & $0.13$ & $0.14$$^{**}$ & $0.28$$^{***}$ & $1.07$ & $0.84$$^{**}$ & $0.76$ & $0.99$ & $0.90$$^{*}$ & $0.82$$^{**}$ \\
Forward-looking trend, attention & $0.11$ & $0.12$ & $0.26$$^{***}$ & $1.07$ & $0.85$$^{**}$ & $0.76$ & $1.00$ & $0.91$$^{*}$ & $0.83$$^{**}$ \\
\bottomrule
\end{tabular}
 }

    \medskip
    \begin{minipage}{0.95\textwidth}
        \footnotesize \textit{Note:} Deep-$\pi^*$ with time-varying volatility, re-estimated every month with the data up to December 2020, \ldots, November 2023; the forward-looking trend with the loss of equation~\eqref{eq:loss_forecast}. One month ahead: the forecast of inflation in the month after the data (36 forecasts, January 2021 to December 2023). Next twelve months: the average of the forecasts of the twelve months after the data, from the same 36 dates; RMSE of average inflation over the twelve months; log score and CRPS averaged over the twelve monthly forecasts. Log-score difference: model minus benchmark. RMSE and CRPS ratios: model over benchmark. Positive differences and ratios below one favour the model. Stars mark Diebold--Mariano p-values below 0.10 ($^{*}$), 0.05 ($^{**}$) and 0.01 ($^{***}$); with 36 forecasts and overlapping averages, the tests are only indicative. Table~\ref{tab:oos_monthly_constant} in Appendix~\ref{app:out_of_sample} reports the estimates with constant volatility.
    \end{minipage}
\end{table}

Over the next twelve months, the RMSE of both trends is 15 to 18\% below that of UC-SV and 24 to 26\% below that of the random walk, and 4 to 7\% above that of the autoregression, a difference that is not significant. The forward-looking trend also gives the most accurate density forecasts of all the models. Its log scores exceed those of the autoregression by 0.11 to 0.13 and those of UC-SV by 0.12 to 0.14, although only the difference of the encoder with UC-SV is significant, and its CRPS is on a par with that of the autoregression. The log scores of the current trend, by contrast, are close to those of both benchmarks over this horizon. The twelve-month object is the one for which the Central Banker's loss is designed, and over the episode the forward-looking trend described the distribution of inflation over the following year better than the current trend did.

\begin{figure}[htbp]
    \centering
    \includegraphics[width=\textwidth]{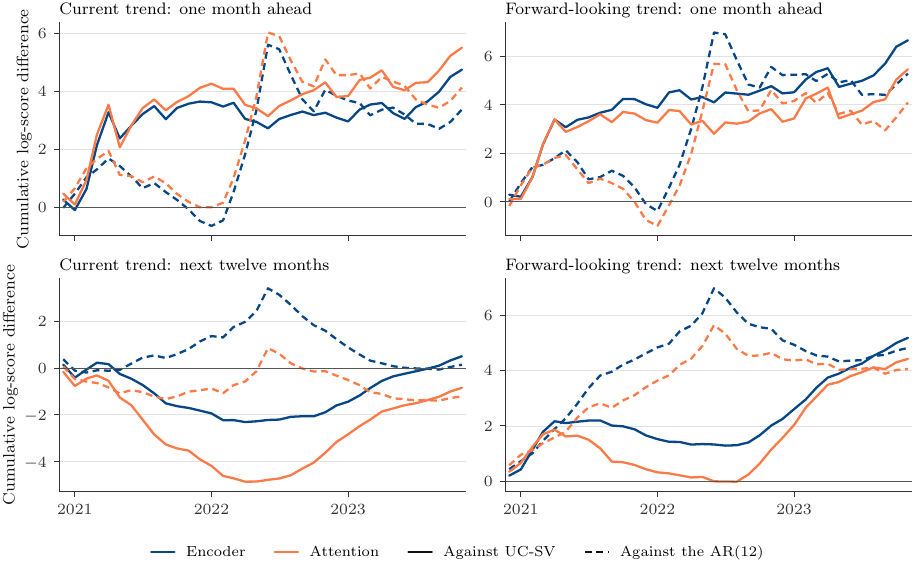}
    \caption{\textbf{Cumulative log-score difference against UC-SV and the autoregression, 2021--2023.} Deep-$\pi^*$ with time-varying volatility, re-estimated every month: the forecasts of inflation one month ahead (top) and over the next twelve months (bottom), of the current trend (left) and of the forward-looking trend (right), for the encoder (blue) and the attention architecture (orange), against UC-SV (solid) and the autoregression (dashed), cumulated over the dates of the forecasts, December 2020 to November 2023. A rising line means that Deep-$\pi^*$ forecasts better than the benchmark over that stretch.}
    \label{fig:oos_cumulative_monthly}
\end{figure}

Figure~\ref{fig:oos_cumulative_monthly} shows when the gains were made. One month ahead, both trends gained against UC-SV in the first months of 2021, when inflation began to rise, and again in the second half of 2023, and against the autoregression from March to June 2022, at the peak of inflation. Over the next twelve months, the forward-looking trend gained against the autoregression during the rise of inflation, from 2021 to the middle of 2022, and gave back part of the gain in the second half of 2022. Against UC-SV, it gained in the first months of 2021, gave back most of this gain by the middle of 2022, and gained again during the fall of inflation, from the autumn of 2022, when the UC-SV trend was still high. The estimates with constant volatility are also more accurate than UC-SV and the random walk in RMSE and CRPS, but by less, and the differences are significant only for the current trend over twelve months (Table~\ref{tab:oos_monthly_constant}).

\subsection{Against Professional Forecasters}\label{subsec:oos_spf}

\begin{figure}[htbp]
    \centering
    \includegraphics[width=\textwidth]{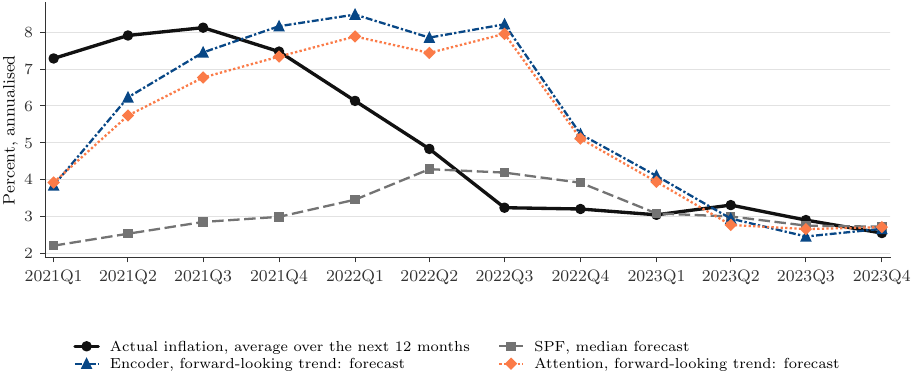}
    \caption{\textbf{Forecasts of average inflation over the next twelve months, 2021--2023.} Each point is one survey, placed at the first month of the survey quarter, 2021Q1 to 2023Q4, the date of the forecasts: actual average inflation over the following twelve months (solid, circles), the median forecast of it in the SPF (dashed, squares), and the forecasts of the forward-looking trend with time-varying volatility, from the estimations with the data up to that month, in the encoder (dash-dotted, triangles) and the attention architecture (dotted, diamonds) (percent, annualised).}
    \label{fig:oos_spf}
\end{figure}

The forecasts of professional forecasters were made with the information available at the time, which makes them a natural benchmark for the forecasts over the next twelve months. The deadline of each quarterly survey of the SPF falls in the middle month of the quarter, so the panellists know the data up to the first month of the quarter, and we compare each survey with the estimations that end in that month. Because the SPF forecasts are for quarters, we convert them to the twelve months after the first month of the quarter by weighting each quarter by its months in that window. Figure~\ref{fig:oos_spf} shows the twelve surveys from 2021 to 2023. Each point refers to one survey: it is placed at the first month of the survey quarter, and it shows the forecasts made in that month of average inflation over the following twelve months, by the SPF and by the forward-looking trend, together with the outcome. The point of July 2022, for instance, compares the forecasts made in July 2022 with average inflation from August 2022 to July 2023. Table~\ref{tab:oos_spf} reports the errors.

During the rise of inflation, Deep-$\pi^*$ was much closer to the outcome than the professional forecasters. In the survey of the first quarter of 2021, the SPF expected inflation of 2.2\% over the next year, against an outcome of 7.3\%, while the forward-looking trend forecast 3.8\% in the encoder and 3.9\% in the attention architecture. By the third quarter of 2021, the SPF still expected 2.8\%, while the forward-looking trend forecast 7.5\% and 6.8\%, against an outcome of 8.1\%. Over the six surveys of the rise, from 2021Q1 to 2022Q2, the SPF forecasts were too low by 3.9 percentage points on average, with an RMSE of 4.3, whereas the forecasts of the forward-looking trend had a bias of less than half a percentage point and an RMSE of 2.2 in both architectures, below that of the autoregression, 2.5.

\begin{table}[htbp]
    \centering
    \small
    \caption{Errors in the forecasts of average inflation over the next twelve months, 2021--2023}
    \label{tab:oos_spf}
    \resizebox{\textwidth}{!}{\begin{tabular}{lrrrrrr}
\toprule
 & \multicolumn{2}{c}{Rise (2021Q1--2022Q2)} & \multicolumn{2}{c}{Fall (2022Q3--2023Q4)} & \multicolumn{2}{c}{2021--2023} \\
\cmidrule(lr){2-3} \cmidrule(lr){4-5} \cmidrule(lr){6-7}
 & Bias & RMSE & Bias & RMSE & Bias & RMSE \\
\midrule
SPF, median forecast & $3.91$ & $4.29$ & $-0.24$ & $0.51$ & $1.84$ & $3.05$ \\
Encoder, forward-looking trend: forecast & $-0.04$ & $2.25$ & $-1.23$ & $2.25$ & $-0.64$ & $2.25$ \\
Attention, forward-looking trend: forecast & $0.45$ & $2.15$ & $-1.15$ & $2.13$ & $-0.35$ & $2.14$ \\
Encoder, current trend: forecast & $1.07$ & $2.42$ & $-1.64$ & $2.48$ & $-0.29$ & $2.45$ \\
Attention, current trend: forecast & $1.13$ & $2.58$ & $-1.21$ & $2.14$ & $-0.04$ & $2.37$ \\
AR(12): forecast & $1.46$ & $2.47$ & $-1.33$ & $1.69$ & $0.06$ & $2.12$ \\
UC-SV: forecast & $1.19$ & $2.82$ & $-1.93$ & $2.53$ & $-0.37$ & $2.68$ \\
Random walk: forecast & $1.66$ & $3.46$ & $-2.46$ & $3.02$ & $-0.40$ & $3.25$ \\
\bottomrule
\end{tabular}
 }

    \medskip
    \begin{minipage}{0.95\textwidth}
        \footnotesize \textit{Note:} Errors: outcome minus forecast (percentage points), so that a positive bias means that the forecasts were too low. Bias: the mean error. Twelve quarterly surveys, 2021Q1 to 2023Q4; each survey is compared with the estimations with the data up to the first month of the survey quarter, and with the benchmarks from the same month. SPF: median forecasts of CPI inflation, converted from quarters to the twelve months after the first month of the quarter. Deep-$\pi^*$: time-varying volatility; forecast: the average of the forecasts of inflation 1 to 12 months ahead.
    \end{minipage}
\end{table}

The professional forecasters anticipated the fall of inflation better. From the third quarter of 2022 their forecasts were close to the outcome, with an RMSE of 0.5, while the forward-looking trend still forecast about 8\% in the third quarter of 2022 and came down over the following three quarters, with an RMSE of 2.1 to 2.3 over the fall. Over the whole episode, the forecasts of the forward-looking trend had an RMSE of 2.1 to 2.2, against 3.1 for the SPF and 2.1 for the autoregression.

 \section{Conclusion}
\label{sec:conclusion}

This paper introduces Deep-$\pi^*$, a model of trend inflation in which the innovations to the trend are a bounded function of a large set of macroeconomic indicators, estimated with neural networks inside a trend-cycle decomposition. The trend keeps the persistence of unobserved components models, and exact group Shapley values trace its movements to groups of indicators. We estimate four architectures, an encoder or a parsimonious similarity-based attention followed by the same encoder, each with constant or time-varying volatility. Two losses define two trends: the current trend, which fits current inflation, and the forward-looking trend, which best predicts inflation over the next year.

For US CPI inflation from 1970 to 2025, the eight estimates agree on the main episodes of US inflation history, and the architecture and the volatility change the trends little. The volatility state follows the stochastic volatility of the UC-SV model and moves more smoothly. The two losses differ in timing and in drivers. In the full-sample estimates, the forward-looking trend peaked eight to twelve months before the current trend after the pandemic. The rise of the current trend came mainly from the labour market. For the forward-looking trend, the encoder points to real activity and the labour market, and the attention architecture to the surge of money and credit in 2020. After the pandemic, the trends moved mainly with indicators of demand and monetary conditions, and they looked through most of the price movements usually attributed to supply shocks. Over the longer history, interest and exchange rates are the largest contributor to the disinflation of the early 1980s, with rising long-term rates lowering the forward-looking trend during the Volcker tightening and again in 2022. The contributions depart clearly from linearity, through both nonlinear effects and interactions between the groups. Re-estimated every month with the data available at the time during the post-pandemic episode, both trends forecast inflation more accurately than the UC-SV model and a random walk and about as accurately as an autoregression, and the forward-looking trend gives the most accurate density forecasts of inflation over the next year. Because the trend in a month depends only on the indicators up to that month, it is revised only when the network is re-estimated: the current trend was revised about as much as the UC-SV trend, and the forward-looking trend mainly in the last months of each sample, where it is still in part a forecast. In real time, the forward-looking trend rose from the start of 2021, when professional forecasters still expected inflation of 3\% or less over the next year, and it turned in 2022 as the contributions of employment and of money and credit declined, although the forecasters anticipated the fall of inflation earlier. The two architectures are about equally accurate in this episode, so the difference between their attributions of the rise of the forward-looking trend reflects model uncertainty.

The two trends answer different questions. The current trend measures the persistent component of inflation today. The forward-looking trend measures the trend inflation today that best predicts inflation over the next year. Reported together with their drivers, they give an account of both the state of underlying inflation and the forces that move its outlook.

The framework opens several avenues for future research. First, the model can be applied to other persistent macroeconomic variables, such as potential output growth or the natural rate of interest, possibly estimated jointly with trend inflation. It could also be applied to the components of inflation, as \citet{StockWatson2016} do with an unobserved components model with stochastic volatility for seventeen components of PCE inflation, whose trends they aggregate into a measure of trend inflation. Second, applying it to euro area data, where the sample is shorter, would test how the architectures perform with fewer observations. Third, following \citet{MontesGaldonOrtega2022}, who let the skewness of structural shocks vary over time, the volatility network could be extended to the asymmetry of inflation risks, so that the model describes more of the distribution of inflation. Finally, the similarity-based causal attention could be applied to other macroeconomic time series in which the samples are too short for standard transformer attention, to see whether it can improve the accuracy of forecasts and the interpretability of the drivers of their movements.

\newpage
\bibliographystyle{apalike}

\newpage
\appendix

\section{Additional Details on the Neural Network Architecture}
\label{app:architecture}

\subsection{Hyperparameter Summary}

Table~\ref{tab:hyperparameters} summarises the key hyperparameters used in the estimation.

\begin{table}[htbp]
\centering
\small
\setlength{\tabcolsep}{4pt}
\caption{Hyperparameter settings}
\label{tab:hyperparameters}
\begin{tabular}{@{}>{\raggedright\arraybackslash}p{6.2cm}>{\raggedright\arraybackslash}p{9.5cm}@{}}
\toprule
\textbf{Parameter} & \textbf{Value} \\
\midrule
\multicolumn{2}{@{}l}{\textit{Model architecture}} \\
\hspace{1em}Encoder layer widths & $122 \to 61 \to 30 \to 15 \to 7 \to 3 \to 1$ \\
\hspace{1em}Activation function & SiLU (Swish) \\
\hspace{1em}Dropout rate & 0.2 (encoder), 0.1 (attention architecture) \\
\hspace{1em}Dropout masks & One per pass, shared by all months \\
\hspace{1em}Weight initialisation & LeCun normal (encoder), Xavier uniform (attention architecture) \\
\hspace{1em}Attention similarity & Cosine, with estimated sharpness $\beta$ \\
\hspace{1em}Attention decay ($\delta$) & Estimated, starting at 0.9 \\
\hspace{1em}Dropout on attention weights & 0.1, a new mask each month \\
\hspace{1em}Residual connection & None \\
\hspace{1em}Layer normalisation & None \\
\midrule
\multicolumn{2}{@{}l}{\textit{Trend dynamics}} \\
\hspace{1em}Persistence ($\theta$) & 0.995 \\
\hspace{1em}Bound on the innovations ($c$) & 40 standardised units (about 155 pp annualised) \\
\hspace{1em}Number of inflation lags ($P$) & 12 \\
\hspace{1em}Sum of AR coefficients & Restricted to $(0, 1)$ \\
\hspace{1em}Forecasting horizon ($H$) & 12 months (Central Banker's loss) \\
\midrule
\multicolumn{2}{@{}l}{\textit{Time-varying volatility}} \\
\hspace{1em}Volatility network & $(k + 12) \to 8 \to 4 \to 1$, dropout 0.1 \\
\hspace{1em}Inputs & Indicators and 12 lagged absolute errors \\
\hspace{1em}Recursion of the state & $\theta = 0.995$, as the trend; $|\lambda_t - \theta\lambda_{t-1}| < \bar{\xi} = 0.1$ \\
\hspace{1em}Range of $\sigma_t$ & $[\bar{\sigma}/5, \, 5\bar{\sigma}]$ ($\bar{v} = \log 5$) \\
\hspace{1em}Horizon factors & $\kappa_2, \ldots, \kappa_{12}$ (Central Banker's loss) \\
\midrule
\multicolumn{2}{@{}l}{\textit{Estimation}} \\
\hspace{1em}Optimiser & AdamW \\
\hspace{1em}Learning rate & 0.001 for the first 36,000 epochs \\
\hspace{1em}Weight decay & 0.1, weight matrices only \\
\hspace{1em}Number of epochs & 60,000 \\
\hspace{1em}Gradient clipping norm & 1.0, all parameters \\
\hspace{1em}Learning-rate schedule & Cosine decay over the last 24,000 epochs \\
\hspace{1em}\quad Final learning rate & $10^{-6}$ \\
\hspace{1em}Likelihood & Student-$t$ ($\nu \approx 3.05$) \\
\hspace{1em}Prior on $\sum_p \phi_p$ & $\mathcal{N}(0.5, 0.2^2)$, weight $1/T$ or $H/N$ \\
\midrule
\multicolumn{2}{@{}l}{\textit{Uncertainty quantification}} \\
\hspace{1em}MC dropout passes & 500 (200 out of sample) \\
\hspace{1em}Band & 80\% (10th--90th percentiles) \\
\hspace{1em}Networks in the final ensemble & 5 \\
\midrule
\multicolumn{2}{@{}l}{\textit{Data}} \\
\hspace{1em}FRED-MD vintage & September 2026 \\
\hspace{1em}Sample & February 1970 to December 2025 ($T = 671$) \\
\hspace{1em}Indicators & 122 \\
\hspace{1em}Standardisation & Median and 2 SD (indicators), SD (inflation) \\
\hspace{1em}Same-month consumer prices & Lagged one month \\
\bottomrule
\end{tabular}
\end{table}

\subsection{Training and Uncertainty Bands}\label{app:training}

Section~\ref{subsec:training} describes the estimation by gradient descent. Each epoch is a single step over the full sample, because the trend in a month depends on the innovations of all earlier months. We use the AdamW optimiser \citep{Loshchilov2019}, a variant of Adam \citep{Kingma2015} that adapts the step size of each parameter and decouples the weight decay from the gradient update, and we train for a fixed number of epochs.

Two further techniques stabilise training. First, we clip the norm of the gradient of all parameters in each step \citep{Pascanu2013}. This guards against occasional large gradients \citep{Bengio1994}, which can arise because, through the recursion in equation~\eqref{eq:state}, the error of each month depends on the innovations of all earlier months, so that the gradient sums the contributions of the whole sample. Second, the learning rate is constant for the first 60\% of the epochs, and then decays to a small value along a cosine curve \citep{LoshchilovHutter2017}. With dropout, each step evaluates a different thinned network, so at a constant learning rate the parameters keep moving around the minimum of the loss. The decay lets them settle there. At the end of training, we check that the trend has stopped changing. Algorithm~\ref{alg:training} summarises the procedure.

\begin{algorithm}[H]
\caption{Training Procedure for Deep-$\pi^*$}\label{alg:training}
\begin{algorithmic}[1]
\Require Indicators $X \in \R^{T \times k}$, inflation $\{\pi_t\}_{t=1}^T$, lags $Z \in \R^{T \times P}$
\Require Fixed values: $\theta$, $c$, $\bar{\xi}$, $\bar{v}$, $\nu$; learning rate $\eta$, weight decay coefficient, epochs $E$, clipping norm $\zeta$
\State Initialise the network weights (LeCun normal for the encoder, Xavier uniform for the encoder of the attention architecture) and biases at zero
\State Initialise $\pi_0^*$, $\sigma$ (or $\bar{\sigma}$, with the last layer of the volatility network at zero) and $\{\phi_p\}_{p=1}^P$
\State Create the AdamW optimiser with learning rate $\eta$ and weight decay, and the learning-rate scheduler
\For{$e = 1, \ldots, E$}
    \State Draw one dropout mask for each layer, shared by all months (and, in the attention architecture, masks on the attention weights that change by month)
    \State $g \gets \text{NeuralNetwork}(X; \Theta)$ \Comment{Forward pass}
    \State $\tau_t \gets c \tanh(g_t / c)$ for $t = 1, \ldots, T$ \Comment{Bounded innovations}
    \For{$t = 1, \ldots, T$} \Comment{Trend recursion}
        \State $\pi_t^* \gets \theta \, \pi_{t-1}^* + (1-\theta) \, \tau_t$
    \EndFor
    \State Compute the residuals $r_t$ (or the forecast errors $r_{t,t+h}$)
    \State With time-varying volatility: compute $\xi_t$ from the inputs $\bm{z}_t$, the state $\lambda_t$ with the recursion \eqref{eq:volstate}, and the scales $\sigma_t$ (or $\sigma_{t,h}$)
    \State $\mathcal{L} \gets \mathcal{L}_0 - \tfrac{1}{T} \log \phi_\mathcal{N}(\sum_p \phi_p \mid \mu_\phi, s_\phi^2)$, or $\mathcal{L}_H$ with weight $\tfrac{H}{N}$ on the prior \Comment{Loss and prior}
    \State Compute $\nabla_\Theta \mathcal{L}$ and clip its norm at $\zeta$
    \State Update $\Theta$ with an AdamW step, and the learning rate with the scheduler
\EndFor
\State \Return Trained model and $\{\pi_t^*\}_{t=1}^T$
\end{algorithmic}
\end{algorithm}

After training, we make $M$ passes with dropout active. In each pass, each layer switches off the same units in every month, so that the pass applies one network to the whole sample and gives one path of the trend. A path must come from a single network because the trend accumulates the innovations of all past months through the recursion in equation~\eqref{eq:state}. Training draws the masks in the same way, so the networks drawn after training are those that training regularised. In the attention architecture, the dropout masks on the attention weights change from month to month, in training as after it, and in the architectures with time-varying volatility each pass also gives a path of the scale.

\subsection{Units and the Bound}\label{app:units}

The model is estimated with inflation in the standardised units of the target: monthly inflation minus its median, divided by its standard deviation, both computed from February 1969 to December 2025, a sample that includes the twelve months that enter only as lags. One unit is 3.9 percentage points of annualised inflation, and zero is the median, 3.3\% annualised. The trend, the innovations and the bound $c$ are in these units, and the figures and tables convert the trend to percent, annualised.

In a steady state the trend equals the innovation, $\bar{\pi}^* = \bar{\tau}$. Trend inflation of 0\% therefore corresponds to an innovation of about $-0.85$, and trend inflation of 12\% to about $2.2$. The innovations of individual months are much larger than these values, because the trend averages them with weights $(1-\theta)\theta^{t-s}$: a move of the trend of 0.25 percentage points in a month requires an innovation about 13 units away from the trend. The bound $c = 40$ lies above the usual size of the innovations, so it compresses most of them little, and it caps the monthly move of the trend at $(1-\theta)(c + |\pi_{t-1}^*|)$, about 0.2 units, or 0.8 percentage points annualised.

In estimation, the output of the last layer of the network is multiplied by a fixed factor of 20. The factor is part of the function $g$ and leaves the model unchanged. It sets the scale of the network output relative to its weights, which matters for their initialisation and for the weight decay.

\subsection{Parameter Count}\label{app:params}

With $k = 122$ indicators and five hidden layers, the encoder has
\[
(122 \times 61 + 61) + (61 \times 30 + 30) + (30 \times 15 + 15) + (15 \times 7 + 7) + (7 \times 3 + 3) + (3 \times 1 + 1) = 9{,}968
\]
parameters, including biases. The attention architecture adds three: the sharpness $\beta$, the decay $\delta$ and the output scale $\gamma$. The twelve autoregressive coefficients are parameterised by their sum, through a logistic function, and by their allocation across lags, which adds 13 parameters. With $\sigma$ and $\pi_0^*$, the encoder model has 9,983 parameters and the attention model 9,986. In the architectures with time-varying volatility, the volatility network adds $(134 \times 8 + 8) + (8 \times 4 + 4) + (4 \times 1 + 1) = 1{,}121$ parameters, $\bar{\sigma}$ takes the place of $\sigma$, and the Central Banker's loss adds the 11 horizon factors $\kappa_2, \ldots, \kappa_{12}$.

These counts are large relative to the sample size ($T = 671$), and several features of the model limit overfitting. First, the network produces a single scalar per month, which constrains its effective capacity. Second, dropout at rates of 10--20\% trains an ensemble of thinned networks \citep{Srivastava2014}. Third, weight decay shrinks the weights towards zero. Fourth, the trend filter in equation~\eqref{eq:state} and the bound on the innovations limit how fast the trend can move, which constrains the effective degrees of freedom further. The out-of-sample evaluation of Section~\ref{sec:out_of_sample} assesses the model on data that it was not estimated on.

\subsection{FRED-MD Variable Groups}\label{app:variable_groups}

Table~\ref{tab:variable_groups} lists the nine groups of indicators used in the decompositions. They are based on the classification of \citet{McCrackenNg2016}, with two differences: real consumption and sales are included in real activity, and four indicators that FRED-MD assigns to money and credit or to the stock market (two consumer-loan series, bank securities and the VIX) form a group of other indicators.

\begin{table}[H]
\centering
\small
\caption{Groups of indicators in the decompositions}
\label{tab:variable_groups}
\begin{tabular}{@{}>{\raggedright\arraybackslash}p{4.6cm}>{\raggedright\arraybackslash}p{9.3cm}r@{}}
\toprule
\textbf{Group} & \textbf{Representative indicators} & \textbf{Count} \\
\midrule
Real activity & Industrial production indices, real personal income, real consumption, capacity utilisation & 19 \\
Employment & Unemployment rate, payroll employment, average hours, initial claims, help-wanted indices & 28 \\
Housing & Housing starts, building permits & 10 \\
Orders and inventories & Manufacturers' new orders, unfilled orders, inventories & 5 \\
Money and credit & M1, M2, real M2, monetary base, reserves, business and real estate loans, consumer credit & 10 \\
Stock market & S\&P 500 index, dividend yield, price-earnings ratio & 3 \\
Interest and exchange rates & Federal funds rate, Treasury yields, commercial paper rates, spreads, exchange rates & 21 \\
Prices & Producer price indices, oil and commodity prices, average hourly earnings, CPI and PCE components (lagged one month) & 22 \\
Others & Consumer motor vehicle loans, total consumer loans and leases, bank securities, VIX & 4 \\
\midrule
Total & & 122 \\
\bottomrule
\end{tabular}
\end{table}

\noindent Note: FRED-MD vintage of September 2026. CPI inflation (CPIAUCSL) is the target and is excluded from the indicators.
 \section{Additional Results on the Trends}
\label{app:volatility_trends}

\begin{figure}[htbp]
    \centering
    \includegraphics[width=\textwidth]{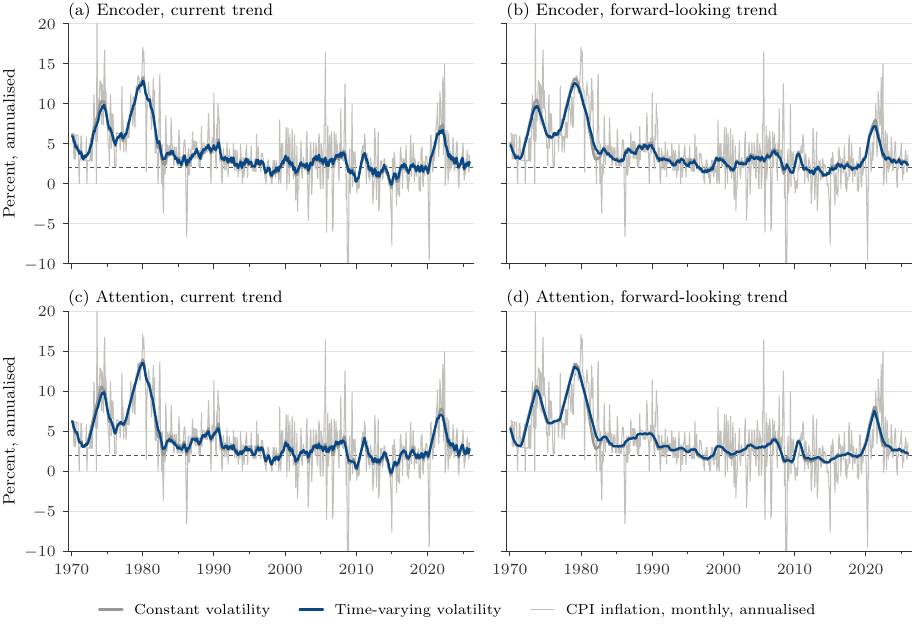}
    \caption{\textbf{The trends with constant and with time-varying volatility.} Deep-$\pi^*$ with constant volatility (grey) and with time-varying volatility (blue), for the encoder (top) and the attention architecture (bottom), under the Econometrician's loss (left) and the Central Banker's loss (right). Monthly CPI inflation, annualised, in light grey. Percent, annualised.}
    \label{fig:trends_volatility}
\end{figure}

\begin{figure}[htbp]
    \centering
    \includegraphics[width=\textwidth]{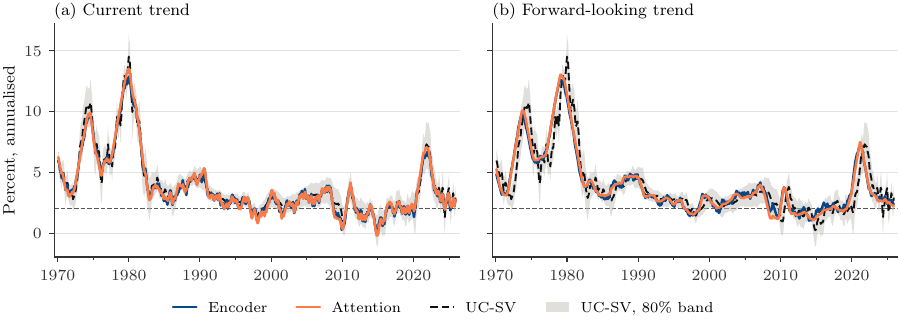}
    \caption{\textbf{The trends and the UC-SV trend.} Deep-$\pi^*$ of the encoder (blue) and of the attention architecture (orange), both with time-varying volatility, and the trend of the UC-SV model of \citet{StockWatson2007} (dashed; posterior mean and 80\% band), estimated on CPI inflation from February 1969 to December 2025 by Gibbs sampling with the precision-based sampler of \citet{ChanJeliazkov2009}: (a) the Econometrician's loss; (b) the Central Banker's loss. Percent, annualised. From February 1970 to December 2025, the correlation between the levels of Deep-$\pi^*$ and of the UC-SV trend is 0.99 for the current trend and 0.90 to 0.92 for the forward-looking trend. The UC-SV trend peaks at 14.5\% in January 1980, and Deep-$\pi^*$ at 12.6\% to 13.5\% between February 1979 and February 1980.}
    \label{fig:trends_ucsv}
\end{figure}

Figure~\ref{fig:trends_volatility} compares the trends of the architectures with constant volatility and with the time-varying volatility of Section~\ref{subsec:volatility}, for both architectures and both losses. Figure~\ref{fig:trends_ucsv} compares the trends with the trend of the UC-SV model of \citet{StockWatson2007}, estimated on monthly CPI inflation from February 1969 to December 2025. These are the inflation data of Deep-$\pi^*$, including the first twelve months, which enter it only as lags. The levels are highly correlated. The UC-SV trend moves about as much from month to month as the current trends and more than the forward-looking trends, and it peaks higher in 1980.

 \section{Turning Points of the Trends}
\label{app:turning_points}

Table~\ref{tab:turning_points} reports the turning points that delimit the episodes of Figure~\ref{fig:episodes}, for the four estimates with time-varying volatility. Each turning point is the highest or lowest value of the trend within a window: the trough between June 1970 and December 1973; the peak between 1978 and 1981; the trough between 1983 and 1987; the peak between 2005 and 2008; the trough between June 2008 and December 2010; the trough between 2019 and 2020; and the peak between June 2020 and June 2023.

\begin{table}[htbp]
    \centering
    \small
    \caption{Turning points of the trends}
    \label{tab:turning_points}
    \begin{tabular}{lrrrr}
\toprule
 & \multicolumn{2}{c}{Current trend} & \multicolumn{2}{c}{Forward-looking trend} \\
\cmidrule(lr){2-3} \cmidrule(lr){4-5}
 & Encoder & Attention & Encoder & Attention \\
\midrule
Trough before the Great Inflation & Sep 1971 & Sep 1971 & Jun 1971 & Jun 1971 \\
Peak of the Great Inflation & Feb 1980 & Jan 1980 & Mar 1979 & Feb 1979 \\
End of the disinflation & Mar 1986 & Mar 1986 & May 1985 & Apr 1985 \\
Peak before the financial crisis & Mar 2008 & Mar 2008 & Mar 2007 & Mar 2007 \\
Trough after the financial crisis & Dec 2009 & Feb 2010 & Oct 2009 & Sep 2009 \\
Trough before the pandemic & Feb 2020 & Jun 2019 & Mar 2019 & Jan 2019 \\
Post-pandemic peak & Mar 2022 & Nov 2021 & Mar 2021 & Mar 2021 \\
End of the sample & Dec 2025 & Dec 2025 & Dec 2025 & Dec 2025 \\
\bottomrule
\end{tabular}
 
    \medskip
    \begin{minipage}{0.95\textwidth}
        \footnotesize \textit{Note:} Months of the turning points of Deep-$\pi^*$, with time-varying volatility. Consecutive turning points delimit the episodes of Figure~\ref{fig:episodes}; the last episode ends in December 2025.
    \end{minipage}
\end{table}
 \section{Robustness of the Drivers}
\label{app:robustness_drivers}

This appendix checks the contributions of the groups of indicators to the rise of the trends from January 2019 to their peaks (Table~\ref{tab:drivers_2019_2022}) against the choices behind them: the networks of each estimate (Table~\ref{tab:robust_networks}), the groups (Table~\ref{tab:robust_finer}), the reference values of the Shapley values (Table~\ref{tab:robust_reference}), and the history of the indicators in the attention architecture (Table~\ref{tab:robust_memory}). The tables use the 40 networks of the final estimates, and no network is re-estimated.

\begin{table}[htbp]
    \centering
    \small
    \caption{The contributions across the networks of each estimate}
    \label{tab:robust_networks}
    \resizebox{\textwidth}{!}{\begin{tabular}{lrrr}
\toprule
 & Employment & Money and credit & Employment + money and credit \\
\midrule
\multicolumn{4}{l}{\textit{Current trend}} \\
Encoder, constant volatility & $2.55$ [$2.05$, $3.13$] & $-0.11$ [$-0.39$, $0.12$] & $2.58$ [$2.05$, $3.16$] \\
Encoder, time-varying volatility & $2.08$ [$1.48$, $2.76$] & $0.49$ [$-0.01$, $1.36$] & $2.62$ [$2.22$, $3.06$] \\
Attention, constant volatility & $3.16$ [$1.59$, $3.90$] & $1.27$ [$0.78$, $2.25$] & $4.10$ [$3.73$, $4.73$] \\
Attention, time-varying volatility & $2.76$ [$2.51$, $3.00$] & $1.03$ [$0.57$, $1.76$] & $3.50$ [$3.10$, $4.20$] \\
\midrule
\multicolumn{4}{l}{\textit{Forward-looking trend}} \\
Encoder, constant volatility & $1.84$ [$1.08$, $2.40$] & $1.76$ [$1.00$, $2.55$] & $3.54$ [$3.13$, $4.24$] \\
Encoder, time-varying volatility & $1.73$ [$1.29$, $2.42$] & $0.66$ [$0.41$, $1.34$] & $2.34$ [$1.78$, $2.91$] \\
Attention, constant volatility & $0.90$ [$0.61$, $1.06$] & $2.65$ [$2.37$, $3.13$] & $3.64$ [$3.19$, $4.21$] \\
Attention, time-varying volatility & $0.81$ [$0.63$, $0.99$] & $2.78$ [$2.55$, $2.96$] & $3.66$ [$3.33$, $4.07$] \\
\bottomrule
\end{tabular}
 }

    \medskip
    \begin{minipage}{0.95\textwidth}
        \footnotesize \textit{Note:} Contributions to the rise of each trend from January 2019 to its peak (exact group Shapley values; percentage points, annualised): the mean over the five networks of each estimate and, in brackets, the lowest and the highest network. Employment + money and credit: the two groups as one group in the Shapley values (eight groups), so its contribution is not exactly the sum of the two.
    \end{minipage}
\end{table}

\begin{table}[htbp]
    \centering
    \small
    \caption{The contributions of finer groups}
    \label{tab:robust_finer}
    \resizebox{\textwidth}{!}{\begin{tabular}{lrrrrr}
\toprule
 & \multicolumn{3}{c}{Employment} & \multicolumn{2}{c}{Money and credit} \\
\cmidrule(lr){2-4} \cmidrule(lr){5-6}
 & Unemployment, claims & Payroll, hours & Vacancies & Money & Credit \\
\midrule
\multicolumn{6}{l}{\textit{Current trend}} \\
Encoder, constant volatility & $-0.88$ & $1.98$ & $1.38$ & $-0.33$ & $0.26$ \\
Encoder, time-varying volatility & $-0.26$ & $1.31$ & $1.01$ & $-0.04$ & $0.41$ \\
Attention, constant volatility & $1.08$ & $1.31$ & $1.22$ & $0.61$ & $0.63$ \\
Attention, time-varying volatility & $1.39$ & $1.09$ & $0.66$ & $0.60$ & $0.43$ \\
\midrule
\multicolumn{6}{l}{\textit{Forward-looking trend}} \\
Encoder, constant volatility & $1.23$ & $0.36$ & $0.14$ & $1.12$ & $0.77$ \\
Encoder, time-varying volatility & $1.27$ & $0.34$ & $0.15$ & $0.78$ & $-0.01$ \\
Attention, constant volatility & $0.58$ & $0.39$ & $-0.09$ & $2.44$ & $0.25$ \\
Attention, time-varying volatility & $0.59$ & $0.29$ & $-0.03$ & $2.45$ & $0.31$ \\
\bottomrule
\end{tabular}
 }

    \medskip
    \begin{minipage}{0.95\textwidth}
        \footnotesize \textit{Note:} Contributions to the rise of each trend from January 2019 to its peak (exact group Shapley values with twelve groups; percentage points, annualised; mean over the five networks of each estimate). Employment is split into unemployment and claims (the unemployment rate and the durations of unemployment, initial claims, and the household survey), payroll employment and hours, and vacancies (the help-wanted indices). Money and credit is split into money (the monetary aggregates, the monetary base and reserves) and credit (business, real estate and consumer loans).
    \end{minipage}
\end{table}

\begin{table}[htbp]
    \centering
    \small
    \caption{The contributions with other reference values}
    \label{tab:robust_reference}
    \resizebox{\textwidth}{!}{\begin{tabular}{lrrrrrrr}
\toprule
 & Real activity & Employment & Housing & Money, credit & Rates & Prices & Other \\
\midrule
\multicolumn{8}{l}{\textit{Current trend}} \\
Encoder, constant volatility & $0.09$ & $0.18$ & $-0.26$ & $0.01$ & $0.04$ & $0.01$ & $0.03$ \\
Encoder, time-varying volatility & $0.07$ & $0.23$ & $-0.37$ & $0.02$ & $0.07$ & $0.00$ & $0.04$ \\
Attention, constant volatility & $-0.15$ & $-0.23$ & $0.30$ & $0.34$ & $-0.20$ & $-0.08$ & $0.04$ \\
Attention, time-varying volatility & $-0.05$ & $0.02$ & $-0.05$ & $0.34$ & $-0.28$ & $-0.02$ & $0.09$ \\
\midrule
\multicolumn{8}{l}{\textit{Forward-looking trend}} \\
Encoder, constant volatility & $0.13$ & $-0.08$ & $-0.12$ & $0.10$ & $-0.05$ & $0.01$ & $0.04$ \\
Encoder, time-varying volatility & $0.07$ & $0.04$ & $-0.17$ & $0.08$ & $-0.04$ & $0.00$ & $0.06$ \\
Attention, constant volatility & $-0.08$ & $-0.04$ & $0.37$ & $-0.40$ & $0.12$ & $-0.03$ & $0.03$ \\
Attention, time-varying volatility & $-0.04$ & $0.14$ & $0.45$ & $-0.52$ & $0.00$ & $-0.01$ & $-0.05$ \\
\bottomrule
\end{tabular}
 }

    \medskip
    \begin{minipage}{0.95\textwidth}
        \footnotesize \textit{Note:} Change in the contributions to the rise of each trend from January 2019 to its peak (percentage points, annualised; mean over the five networks of each estimate) when the reference values of the Shapley values are the 2010--2019 averages of the indicators instead of their medians. Money, credit: money and credit. Rates: interest and exchange rates.
    \end{minipage}
\end{table}

\begin{table}[htbp]
    \centering
    \small
    \caption{The memory of the attention architecture}
    \label{tab:robust_memory}
    \resizebox{\textwidth}{!}{\begin{tabular}{lrrrr}
\toprule
 & \multicolumn{2}{c}{Employment} & \multicolumn{2}{c}{Money and credit} \\
\cmidrule(lr){2-3} \cmidrule(lr){4-5}
 & From past months & Shapley value & From past months & Shapley value \\
\midrule
Attention, constant volatility & $0.33$ [$0.25$, $0.43$] & $0.90$ [$0.61$, $1.06$] & $1.06$ [$0.76$, $1.76$] & $2.65$ [$2.37$, $3.13$] \\
Attention, time-varying volatility & $0.36$ [$0.30$, $0.44$] & $0.81$ [$0.63$, $0.99$] & $1.47$ [$1.05$, $1.90$] & $2.78$ [$2.55$, $2.96$] \\
\bottomrule
\end{tabular}
 }

    \medskip
    \begin{minipage}{0.95\textwidth}
        \footnotesize \textit{Note:} Forward-looking trend, attention architecture. From past months: the part of a group's contribution to the rise of the trend from January 2019 to its peak that comes from the past months of its indicators. With the attention weights fixed, it is the trend minus the trend when only the current month of the group's indicators enters the weighted average. Shapley value: the group's contribution in Table~\ref{tab:drivers_2019_2022}. Mean over the five networks and, in brackets, the lowest and the highest network (percentage points, annualised).
    \end{minipage}
\end{table}
 \section{A Linear Encoder}
\label{app:linear}

To measure what the neural network adds, we estimate Deep-$\pi^*$ with a linear encoder. The network output is then a linear function of the indicators, $g(\bx_t) = \bm{w}'\bx_t + b$, and the innovation is its bounded transformation in equation~\eqref{eq:innovation}. The bound, the recursion of the trend, the losses, the data and the training are those of the encoder with constant volatility. Without hidden layers, the linear encoder also has no dropout, so the comparison measures what the network adds with its nonlinearity and its regularisation together. As for the final estimates, each loss averages five networks estimated on the full sample, and out of sample we re-estimate one network every month over 2021--2023, as in Section~\ref{sec:out_of_sample}.

\begin{figure}[htbp]
    \centering
    \includegraphics[width=\textwidth]{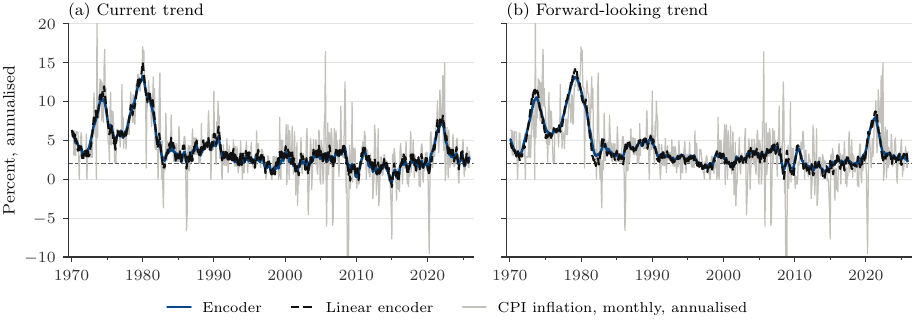}
    \caption{\textbf{The trends of the linear encoder.} Deep-$\pi^*$ of the encoder (blue) and of the linear encoder (dashed), both with constant volatility: (a) the Econometrician's loss; (b) the Central Banker's loss. Each trend is the average of five networks. Monthly CPI inflation, annualised, in grey; the dashed line marks 2\%. Percent, annualised. The vertical axis cuts off three months of inflation: August 1973, and October and November 2008.}
    \label{fig:linear_trends}
\end{figure}

Figure~\ref{fig:linear_trends} compares the trends. The linear trend follows the same path as that of the encoder, with correlations of 0.98 and 0.99 between the levels, but it is much noisier. The standard deviation of its monthly changes is 0.68 percentage points for the current trend and 0.48 for the forward-looking trend, against 0.32 and 0.22 for the encoder (Table~\ref{tab:linear_full_sample}). The linear trend also peaks higher in 1979--80 and in 2021--2022.

\begin{table}[htbp]
    \centering
    \small
    \caption{The trends of the linear encoder and of the encoder, full sample}
    \label{tab:linear_full_sample}
    \begin{tabular}{lrrr}
\toprule
 & \multicolumn{1}{c}{Correlation} & \multicolumn{2}{c}{SD of monthly changes} \\
\cmidrule(lr){2-2} \cmidrule(lr){3-4}
 & of levels & Linear & Encoder \\
\midrule
Current trend & $0.98$ & $0.68$ & $0.32$ \\
Forward-looking trend & $0.99$ & $0.48$ & $0.22$ \\
\bottomrule
\end{tabular}
 
    \medskip
    \begin{minipage}{0.95\textwidth}
        \footnotesize \textit{Note:} Constant volatility. Correlation of levels: between the trends of the linear encoder and of the encoder, February 1970 to December 2025. SD of monthly changes: standard deviation of the monthly changes of each trend (percentage points).
    \end{minipage}
\end{table}

Out of sample, in the post-pandemic episode, the linear encoder is less accurate (Table~\ref{tab:linear_comparison}). The nonlinear network matters mostly for the current trend, whose log scores are higher than those of the linear encoder by 0.22 one month ahead and 0.54 over twelve months, both significant. For the forward-looking trend, the two encoders give point forecasts of similar accuracy in that episode, and the log scores of the encoder are higher by 0.07 and 0.11, differences that are not significant.

\begin{table}[htbp]
    \centering
    \small
    \caption{The linear encoder out of sample}
    \label{tab:linear_comparison}
    \resizebox{\textwidth}{!}{\begin{tabular}{lrrrrrr}
\toprule
 & \multicolumn{3}{c}{Log-score difference, linear $-$} & \multicolumn{3}{c}{RMSE ratio, linear /} \\
\cmidrule(lr){2-4} \cmidrule(lr){5-7}
 & Encoder & AR(12) & UC-SV & Encoder & AR(12) & UC-SV \\
\midrule
Current trend: one month ahead & $-0.22$$^{**}$ & $-0.20$$^{*}$ & $-0.16$ & $1.02$ & $1.02$ & $0.97$ \\
Current trend: next twelve months & $-0.54$$^{***}$ & $-0.53$$^{**}$ & $-0.52$ & $1.14$ & $1.24$ & $0.98$ \\
Forward-looking trend: one month ahead & $-0.07$ & $0.00$ & $0.04$ & $1.04$ & $1.04$ & $0.99$ \\
Forward-looking trend: next twelve months & $-0.11$ & $-0.04$ & $-0.03$ & $1.00$ & $1.09$ & $0.86$ \\
\bottomrule
\end{tabular}
 }

    \medskip
    \begin{minipage}{0.95\textwidth}
        \footnotesize \textit{Note:} Constant volatility. The forecasts of inflation one month ahead and over the next twelve months, as in Table~\ref{tab:oos_monthly}: monthly re-estimation, December 2020 to November 2023. Log-score difference and RMSE ratio: the linear encoder minus, or over, the model in the column. Stars mark Diebold--Mariano p-values below 0.10 ($^{*}$), 0.05 ($^{**}$) and 0.01 ($^{***}$).
    \end{minipage}
\end{table}

Employment contributes most to the rise of both trends of the linear encoder, 4.02 and 4.42 percentage points, as it does, by less, to both trends of the encoder (Table~\ref{tab:linear_drivers}). The linear map, however, spreads the rise over larger contributions of opposite signs. Interest and exchange rates add 3.43 points to the current trend and prices subtract 1.58, and housing subtracts 3.68 points from the forward-looking trend. With its nonlinearity and its regularisation, the network gives a smoother trend, more accurate predictive densities, and contributions of moderate size.

\begin{table}[htbp]
    \centering
    \small
    \caption{The drivers of the rise since 2019: the linear encoder and the encoder}
    \label{tab:linear_drivers}
    \begin{tabular}{lrrrr}
\toprule
 & \multicolumn{2}{c}{Current trend} & \multicolumn{2}{c}{Forward-looking trend} \\
\cmidrule(lr){2-3} \cmidrule(lr){4-5}
 & Linear & Encoder & Linear & Encoder \\
\midrule
Real activity & $1.08$ & $1.84$ & $2.33$ & $1.40$ \\
Employment & $4.02$ & $2.55$ & $4.42$ & $1.84$ \\
Housing & $-0.03$ & $0.48$ & $-3.68$ & $-0.41$ \\
Money and credit & $0.88$ & $-0.11$ & $2.15$ & $1.76$ \\
Interest and exchange rates & $3.43$ & $0.46$ & $1.66$ & $0.98$ \\
Prices & $-1.58$ & $0.04$ & $0.08$ & $0.31$ \\
Other & $-1.51$ & $0.34$ & $-0.74$ & $-0.28$ \\
Total & $6.44$ & $5.77$ & $6.11$ & $5.54$ \\
\bottomrule
\end{tabular}
 
    \medskip
    \begin{minipage}{0.95\textwidth}
        \footnotesize \textit{Note:} Contributions of the groups of indicators to the rise of each trend from January 2019 to its peak (exact group Shapley values; percentage points, annualised; average over five networks), with constant volatility. The peaks are in March 2022 for the current trend of both, and in July 2021 (linear encoder) and June 2021 (encoder) for the forward-looking trend. Total: the change in the trend, which also includes the small change of the path with all indicators at their medians. Other: orders and inventories, the stock market and the other indicators.
    \end{minipage}
\end{table}
 \clearpage
\section{Out-of-Sample Evaluation: Additional Results}
\label{app:out_of_sample}

This appendix complements Section~\ref{sec:out_of_sample} with the results of the post-pandemic exercise that the main text leaves out: the estimates with constant volatility, the peaks of the trends in real time, and the comparison of the two architectures.

\begin{figure}[htbp]
    \centering
    \includegraphics[width=\textwidth]{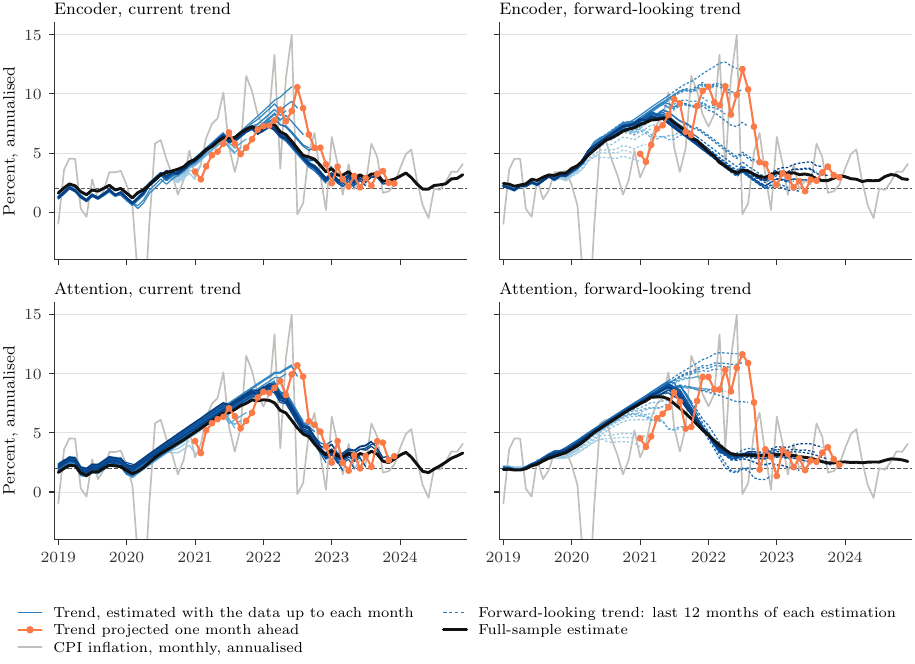}
    \caption{\textbf{The trend re-estimated every month, constant volatility.} As Figure~\ref{fig:oos_spaghetti}, for the architectures with constant volatility.}
    \label{fig:oos_spaghetti_constant}
\end{figure}

Table~\ref{tab:oos_monthly_constant} repeats Table~\ref{tab:oos_monthly} for the estimates with constant volatility, and Figure~\ref{fig:oos_spaghetti_constant} repeats Figure~\ref{fig:oos_spaghetti}. Table~\ref{tab:oos_peaks} compares the peaks of the trends in real time with those of the full-sample estimates, in all eight estimates. Table~\ref{tab:oos_encoder_attention} compares the two architectures, and Table~\ref{tab:oos_pool} compares an equal-weight pool of the two with each of them and with the benchmarks.

\begin{table}[htbp]
    \centering
    \small
    \caption{Forecast accuracy against the benchmarks, 2021--2023: constant volatility}
    \label{tab:oos_monthly_constant}
    \resizebox{\textwidth}{!}{\begin{tabular}{lrrrrrrrrr}
\toprule
 & \multicolumn{3}{c}{Log-score difference} & \multicolumn{3}{c}{RMSE ratio} & \multicolumn{3}{c}{CRPS ratio} \\
\cmidrule(lr){2-4} \cmidrule(lr){5-7} \cmidrule(lr){8-10}
 & AR(12) & UC-SV & Random walk & AR(12) & UC-SV & Random walk & AR(12) & UC-SV & Random walk \\
\midrule
\multicolumn{10}{l}{\textit{One month ahead}} \\
Current trend, encoder & $0.03$ & $0.06$ & $0.18$ & $1.01$ & $0.95$ & $0.92$ & $0.98$ & $0.91$ & $0.85$ \\
Current trend, attention & $0.00$ & $0.04$ & $0.15$ & $0.99$ & $0.93$ & $0.91$ & $0.98$ & $0.91$ & $0.85$ \\
Forward-looking trend, encoder & $0.07$ & $0.11$ & $0.22$ & $1.00$ & $0.95$ & $0.92$ & $0.96$ & $0.89$ & $0.83$ \\
Forward-looking trend, attention & $0.03$ & $0.07$ & $0.18$ & $1.02$ & $0.96$ & $0.93$ & $0.98$ & $0.92$ & $0.85$ \\
\midrule
\multicolumn{10}{l}{\textit{Next twelve months}} \\
Current trend, encoder & $0.01$ & $0.02$ & $0.16$ & $1.08$ & $0.86$$^{***}$ & $0.77$$^{**}$ & $1.03$ & $0.94$ & $0.86$$^{*}$ \\
Current trend, attention & $-0.10$ & $-0.09$ & $0.05$ & $1.12$ & $0.88$$^{***}$ & $0.79$$^{*}$ & $1.06$ & $0.97$ & $0.88$ \\
Forward-looking trend, encoder & $0.08$ & $0.09$ & $0.22$ & $1.09$ & $0.86$ & $0.77$ & $1.01$ & $0.92$ & $0.84$ \\
Forward-looking trend, attention & $-0.04$ & $-0.03$ & $0.10$ & $1.15$ & $0.91$ & $0.81$ & $1.06$ & $0.96$ & $0.88$ \\
\bottomrule
\end{tabular}
 }

    \medskip
    \begin{minipage}{0.95\textwidth}
        \footnotesize \textit{Note:} As Table~\ref{tab:oos_monthly}, for the estimates with constant volatility.
    \end{minipage}
\end{table}

\begin{table}[htbp]
    \centering
    \small
    \caption{The trends in real time and in the full sample, 2020--2023}
    \label{tab:oos_peaks}
    \begin{tabular}{lrrr}
\toprule
 & \multicolumn{2}{c}{Peak, 2020--2023} & \multicolumn{1}{c}{Mean absolute} \\
\cmidrule(lr){2-3} \cmidrule(lr){4-4}
 & Real time & Full sample & revision \\
\midrule
\multicolumn{4}{l}{\textit{Current trend}} \\
Encoder, time-varying volatility & 9.4 (Jun 2022) & 6.7 (Mar 2022) & 1.0 \\
Encoder, constant volatility & 10.6 (Jun 2022) & 7.4 (Mar 2022) & 0.9 \\
Attention, time-varying volatility & 10.0 (Jun 2022) & 7.1 (Nov 2021) & 1.2 \\
Attention, constant volatility & 10.7 (Jun 2022) & 7.8 (Jan 2022) & 1.0 \\
\midrule
\multicolumn{4}{l}{\textit{Forward-looking trend}} \\
Encoder, time-varying volatility & 10.7 (Jun 2022) & 7.2 (Mar 2021) & 2.3 \\
Encoder, constant volatility & 12.1 (Jun 2022) & 8.0 (Jun 2021) & 2.2 \\
Attention, time-varying volatility & 10.7 (Jun 2022) & 7.5 (Mar 2021) & 2.1 \\
Attention, constant volatility & 11.7 (Jun 2022) & 8.0 (Apr 2021) & 2.3 \\
\midrule
\multicolumn{4}{l}{\textit{Benchmark}} \\
UC-SV & 10.7 (Mar 2022) & 7.3 (Nov 2021) & 1.2 \\
\bottomrule
\end{tabular}
 
    \medskip
    \begin{minipage}{0.95\textwidth}
        \footnotesize \textit{Note:} Monthly re-estimation, December 2020 to November 2023 (percent, annualised). Real time: the highest value of the trend in the last month of each estimation, and its month. Full sample: the highest value of the full-sample estimate between January 2020 and December 2023, and its month. Mean absolute revision: the average absolute difference between the trend in real time and the full-sample estimate of the same month, over the 36 months. UC-SV: the trend of the UC-SV model of \citet{StockWatson2007}, estimated at each month with the data up to that month (its filtered estimate), against its estimate with the full sample (its smoothed estimate).
    \end{minipage}
\end{table}

\begin{table}[htbp]
    \centering
    \small
    \caption{The attention architecture against the encoder, 2021--2023}
    \label{tab:oos_encoder_attention}
    \begin{tabular}{lrr}
\toprule
 & \multicolumn{2}{c}{Attention against the encoder} \\
\cmidrule(lr){2-3}
 & Log-score difference & RMSE ratio \\
\midrule
\multicolumn{3}{l}{\textit{One month ahead}} \\
Current trend, time-varying volatility & $0.02$ & $0.98$ \\
Current trend, constant volatility & $-0.02$ & $0.98$ \\
Forward-looking trend, time-varying volatility & $-0.03$ & $1.02$$^{*}$ \\
Forward-looking trend, constant volatility & $-0.04$ & $1.01$ \\
\midrule
\multicolumn{3}{l}{\textit{Next twelve months}} \\
Current trend, time-varying volatility & $-0.04$ & $0.99$ \\
Current trend, constant volatility & $-0.11$$^{*}$ & $1.03$ \\
Forward-looking trend, time-varying volatility & $-0.02$ & $1.01$ \\
Forward-looking trend, constant volatility & $-0.12$$^{**}$ & $1.05$$^{**}$ \\
\bottomrule
\end{tabular}
 
    \medskip
    \begin{minipage}{0.95\textwidth}
        \footnotesize \textit{Note:} The forecasts of inflation one month ahead (top) and over the next twelve months (bottom), from the monthly re-estimation of Section~\ref{sec:out_of_sample}, December 2020 to November 2023. Log-score difference: attention architecture minus encoder. RMSE ratio: attention architecture over encoder. Positive differences and ratios below one favour the attention architecture. Stars mark Diebold--Mariano p-values below 0.10 ($^{*}$), 0.05 ($^{**}$) and 0.01 ($^{***}$).
    \end{minipage}
\end{table}

\begin{table}[htbp]
    \centering
    \small
    \caption{An equal-weight pool of the two architectures, 2021--2023}
    \label{tab:oos_pool}
    \resizebox{\textwidth}{!}{\begin{tabular}{lrrrrrrrr}
\toprule
 & \multicolumn{4}{c}{Log-score difference, pool $-$} & \multicolumn{4}{c}{RMSE ratio, pool /} \\
\cmidrule(lr){2-5} \cmidrule(lr){6-9}
 & Encoder & Attention & AR(12) & UC-SV & Encoder & Attention & AR(12) & UC-SV \\
\midrule
Current trend: one month ahead & $0.01$ & $-0.01$ & $0.11$ & $0.15$$^{**}$ & $0.99$$^{*}$ & $1.01$ & $0.97$ & $0.92$$^{**}$ \\
Current trend: next twelve months & $-0.01$ & $0.03$ & $-0.01$ & $0.00$ & $0.99$ & $1.00$ & $1.04$ & $0.83$$^{***}$ \\
Forward-looking trend: one month ahead & $-0.01$ & $0.02$$^{*}$ & $0.13$ & $0.17$$^{**}$ & $1.01$$^{*}$ & $0.99$$^{**}$ & $0.99$ & $0.94$ \\
Forward-looking trend: next twelve months & $0.00$ & $0.02$ & $0.13$ & $0.14$$^{*}$ & $1.00$ & $0.99$ & $1.07$ & $0.84$$^{**}$ \\
\bottomrule
\end{tabular}
 }

    \medskip
    \begin{minipage}{0.95\textwidth}
        \footnotesize \textit{Note:} Monthly re-estimation, December 2020 to November 2023. Pool: the equal-weight mixture of the encoder and the attention architecture with the same loss, with time-varying volatility. Its log score for each forecast is the log of the average of the two densities, and its point forecast is the average of their point forecasts. Log-score difference: pool minus the model in the column. RMSE ratio: pool over the model in the column. Stars mark Diebold--Mariano p-values below 0.10 ($^{*}$), 0.05 ($^{**}$) and 0.01 ($^{***}$).
    \end{minipage}
\end{table}

\end{document}